\documentclass[twocolumn]{aa} 

\usepackage{graphicx}
\usepackage{txfonts}
\usepackage[linkcolor=black, bookmarksopen=true, colorlinks=true, citecolor=black]{hyperref}
\usepackage{float}
\usepackage{lscape}

\makeatletter
\def\@biblabel#1{}
\makeatother
\renewcommand{\cite}{\citealp}

\begin{document}

   \title{New spectroscopic binary companions of giant stars and updated metallicity distribution for binary systems
 \thanks{Based on observations collected at La Silla - Paranal Observatory under
programs IDs IDs 085.C-0557, 087.C.0476, 089.C-0524, 090.C-0345, 096.A-9020 and through the Chilean Telescope Time under programs
IDs CN2012A-73, CN2012B-47, CN2013A-111, CN2013B-51, CN2014A-52 and CN2015A-48.}}

      
   \author{P.\,Bluhm \inst{1}
         \and
          M.I.\,Jones \inst{2}
         \and
          L.\,Vanzi   \inst{2,}\inst{3}
         \and
          M.G.\,Soto \inst{4}
         \and
         J. Vos \inst{5}
         \and
          R.A.\,Wittenmyer \inst{6,}\inst{7}
         \and
          H.\,Drass  \inst{2}
         \and
         J.S.\,Jenkins \inst{4}
         \and
          F.\,Olivares \inst{8,}\inst{9}
         \and
          R.E.\,Mennickent \inst{10}
         \and
          M.\,Vu\v{c}kovi\'c \inst{5}
         \and
          P.\,Rojo \inst{4}
         \and
          C.H.F.\,Melo \inst{11}
          }
          
   \offprints{P.Bluhm, \email pvbluhm@uc.cl}

    \institute{Instituto de Astrof\'isica, Pontificia Universidad Cat\'olica,
               Av.Vicu\~na Mackenna 4860, 782-0436 Macul, Santiago, Chile.
         \and
             Center of Astro-Engineering UC, Pontificia Universidad Cat\'olica, Santiago, Chile.
         \and
             Department of Electrical Engineering, Pontificia Universidad Cat\'olica, Santiago, Chile.
         \and
             Departamento de Astronom\'ia,
             Universidad de Chile, Camino El Observatorio 1515, Las Condes, Santiago, Chile.
         \and
             Instituto de F\'isica y Astronom\'ia,
             Universidad de Vapara\'iso, Valpara\'iso, Chile.
        \and
             School of Physics and Australian Centre for Astrobiology, University of New South Wales,
             Sydney 2052, Australia.
         \and
             Computational Engineering and Science Research Centre, University of Southern
             Queensland, Toowoomba, Australia.    
          \and
             Departamento de Ciencias Fisicas,
             Universidad Andres Bello, Avda. Republica 252, Santiago, Chile.
          \and
             Millennium Institute of Astrophysics, Santiago, Chile.
         \and
             Departamento de Astronom\'ia,
             Universidad de Concepci\'on, Casilla 160-C Concepci\'on, Chile.                      
         \and
              European Southern Observatory,
              Casilla 19001, Santiago, Chile.
             }

   \date{}

\abstract
{We report the discovery of 24 spectroscopic binary companions to giant stars. 
We fully constrain the orbital solution for 6 of these systems. We cannot unambiguously derive the 
orbital elements for the remaining stars because the phase coverage is incomplete. Of these stars, 6 present radial velocity trends 
that are compatible with long-period brown dwarf companions.\newline\indent
The orbital solutions of the 24 binary systems indicate that these giant binary systems have a wide range in orbital periods,
eccentricities, and companion masses.
 For the binaries with restricted orbital solutions, we find a range of orbital periods of
between $\sim$\,97-1600 days and eccentricities of between $\sim$\,0.1-0.4. \newline\indent
In addition, we studied the metallicity distribution of single and binary giant stars. We computed the metallicity of a total of 
395 evolved stars, 59 of wich are in binary systems. We find a flat distribution for these binary stars and therefore 
conclude that stellar binary systems, and potentially brown dwarfs, have a different formation mechanism than planets.
This result is confirmed by recent works showing that extrasolar planets orbiting giants are more frequent around metal-rich stars.\newline\indent
Finally, we investigate the eccentricity as a function of the orbital period. We analyzed a total of 130 spectroscopic binaries, 
including those presented here and systems from the literature. We find that most of the binary stars with periods $\lesssim$ 30 
days have circular orbits, while at longer orbital periods we observe a wide spread in their eccentricities.}


 
\titlerunning{Spectroscopic binary companions around giant stars}

   \maketitle
%

\section{Introduction}

The study of stars in binary systems provides valuable information about the formation and dynamical
evolution of stars. Radial velocity (RV) surveys have revealed that a significant fraction of the stars in the 
solar neighborhood are found in multiple systems. Duquennoy et al.\,(\cite{DUQ91}) showed that
more than half of the nearby stars are found in multiple systems, although more recent results show that this 
fraction is slightly lower (Lada\,\cite{LAD06}; Raghavan et al.\,\cite{RAG10}).\newline\indent
It is well known that stellar systems predominantly form through the gravitational collapse of the molecular cloud, while
planetary systems are subsequently formed in the protoplanetary disk.
Machida\,(\cite{machida}) investigated the evolution of clouds with various metallicities
and showed that the binary frequency increases as the metallicity decreases.
On the other hand, the planetary formation follows the planet–metallicity correlation. This correlation tells us 
that planets form more efficiently around metal-rich stars (Gonzalez, \cite{GON97}; Santos et al. \cite{SAN01}).
When one of the stars in older stellar systems evolves off of the main sequence, the mutual effect of 
tidal interaction between them might dictate the final orbital configuration of the system.
Verbunt\,\&\,Phinney\,(\cite{VER95}) studied the orbital properties of binaries containing giant stars in open clusters. 
They showed that most of the binaries with periods shorter than $\sim$\,200 days present nearly circular orbits, which is most 
likely explained by the effect of the tidal circularization
(Zahn\,\cite{ZAN77},\,\cite{ZAN89};\,Tassoul\,\cite{Tas87},\,\cite{Tas88},\,\cite{Tas92}).
Similarly, Pan\,et\,al.\,(\cite{Pan98}) showed that the predictions of Zahn's theories on synchronization 
for main-sequence binary systems are compatible with observational data.
In addition, Massarotti\,et\,al.\,(\cite{Massarotti08},\,MAS08 hereafter) showed based on a sample 761 giant stars that all stars in
binary systems with periods shorter than 20 days have circularized orbits. 
They also demonstrated that $\sim$50$\%$ of the orbits that have periods in the range of 20–100 days 
show significant eccentricity. This result shows the importance of studying the 
eccentricity distribution of binary systems containing giant stars. This allows us to test the validity of 
the tidal dissipation theory and to empirically measure the tidal dissipation efficiency. Moreover, these results can be also used  
to study the orbital evolution of planetary systems around evolved stars (e.g., Sato\,et\,al.\,\cite{SAT08}; Villaver\,\&\,Livio\,\cite{VIL09}). 
\newline\indent
In this paper we report the discovery of 24 spectroscopic binary companions to giant stars, wich have been targeted since 2009 by the EXPRESS 
project ({\bf EX}o{\bf P}lanets a{\bf R}ound {\bf E}volved {\bf S}tar{\bf S}; Jones\,et\,al.\,\cite{JON11}).
The parent sample comprises 166 relatively bright giant stars.
The RV measurements of these stars have revealed large amplitude variations, which are explained by the
Doppler shift induced by stellar companions or massive brown dwarfs. For six of them, we have good phase coverage, thus the orbital solution is well 
constrained.
The remaining systems present much longer orbital periods, wich means that either their orbital solution is degenerate,
or they present a linear RV trend.\newline\indent
In addition, we study the metallicity distribution of binary giant stars. To do so, we added 232 giant stars
to the original sample, giving a total of 395 giant stars.
We also investigated the period-eccentricity relation for 130 spectroscopic binary giant stars to understand the role
 of tidal circularization in these systems. \newline\indent
The paper is organized as follows: in Sect.\,\ref{sec2} we briefly describe the observations and data reduction analysis. 
In Sect.\,\ref{sec3} we present the stellar properties
of the primary star. In Sect.\,\ref{sec4} we present the orbital parameters of the binary companions. Finally, 
in Sect.\,\ref{sec5.1} we present the metallicity distribution for the binary system
fraction in giant stars, and in Sect.\,\ref{sec5.2} we present a statistical analysis for the eccentricity distribution.

\begin{table}[]
\centering
  \caption{Instrument descriptions.}
  \label{tb1:Instruments}
\begin{tabular}{llccc}
\hline\hline
Instrument  & Resolution & Range & Exp. time \\ 
            &            & (\AA) & (sec)   &      \\
\hline \vspace{-0.3cm} \\
FEROS       &48000      &3500-9200    &  60-500    \\ 
FECH        &43000      &4000-7000    &  300-600   \\
CHIRON      &80000      &4100-8700    &  500-1000  \\ 
PUCHEROS    &20000      &4000-7000    &  900-1200  \\ 
UCLES       &45000      &3000-7000    &  300-1200  \\ 
HARPS       &115000     &3800-6700    &  90        \\ 
\hline \vspace{-0.3cm} \\
\hline\hline
\end{tabular}
\end{table}


\begin{table*}[t]
\centering
  \caption{Stellar parameters of the primary stars. Error on T$_{\rm{eff}}$ is 100 K.}
  \label{tb2:stellar}
\begin{tabular}{rcccccccc}
\hline\hline
HIP   & B-V        &  V    & T$_{{\rm eff}}$ &$\log{g}$       & L$_\star$   & M$_\star$   &Distance & R$_\star$  \\
      &(mag)       & (mag) & (K)             &(cm\,s$^{-2})$ & (L$_\odot$) & (M$_\odot)$ & (pc)    & (R$_\odot$)  \\
\hline \vspace{-0.3cm} \\
4618  &1.08 (0.013)& 7.79 (0.009)& 4750 &2.91  &19.68 (3.68) &1.47 (0.26) &142.2 (11.5)&\,\,\,6.6  (0.8) \\
7118  &1.06 (0.005)& 5.80 (0.003)& 4820 &2.74  &60.67 (7.60) &1.75 (0.46) &102.4 (4.3) &11.3 (0.9) \\
10548 &0.96 (0.011)& 7.32 (0.008)& 4980 &3.36  &11.10 (1.42) &1.66 (0.10) &86.4 (3.8)  &\,\,\,4.5  (0.4)  \\
22479 &0.99 (0.003)& 5.03 (0.002)& 4990 &2.93  &61.66 (6.32) &2.58 (0.19) &72.3 (1.6)  &10.7  (0.9)\\
59016 &1.06 (0.008)& 7.05 (0.006)& 4800 &2.88  &19.66 (2.81) &1.62 (0.23) &102.6 (5.6) &\,\,\,6.4   (0.6)\\
59367 &1.05 (0.021)& 8.05 (0.014)& 4960 &3.08  &10.47 (1.98) &1.52 (0.15) &99.4 (8.2)  &\,\,\,4.4   (0.5)\\
64647 &1.09 (0.018)& 7.83 (0.012)& 4870 &2.92  &22.01 (4.73) &1.72 (0.27) &149.3 (14.5)&\,\,\,6.7   (0.8)\\
64803 &0.94 (0.006)& 5.12 (0.003)& 5060 &2.63  &67.11 (6.91) &2.71 (0.19) &79.0 (1.7)  &10.8  (0.8)\\
66924 &1.00 (0.006)& 5.96 (0.004)& 4860 &2.53  &63.85 (8.38) &1.72 (0.34) &110.4 (5.1) &11.3  (0.9)\\
67890 &1.14 (0.007)& 6.05 (0.005)& 4750 &2.81  &20.55 (2.23) &1.75 (0.20) &64.9 (1.8)  &\,\,\,6.8   (0.6)\\
68099 &0.95 (0.009)& 6.83 (0.007)& 5130 &3.00  &\,\,69.42 (14.36) &2.87 (0.18) &168.1 (15.5)&10.7  (1.3)\\
71778 &0.95 (0.020)& 7.88 (0.014)& 5040 &3.45  &\,7.24 (1.23) &1.47 (0.13) &95.1 (6.8)  &\,\,\,3.5   (0.4)\\
73758 &1.17 (0.019)& 7.90 (0.012)& 4840 &3.20  &\,5.43 (0.80) &1.36 (0.10) &82.2 (4.7)  &\,\,\,3.4   (0.3)\\
74188 &1.05 (0.014)& 7.13 (0.010)& 4750 &2.95  &12.20 (1.83) &1.36 (0.21) &80.3 (4.7)  &\,\,\,5.2   (0.5)\\
75331 &1.10 (0.015)& 7.59 (0.010)& 4880 &3.33  &\,\,4.97 (0.62) &1.35 (0.10) &66.3 (2.7)  &\,\,\,3.1   (0.3)\\
76532 &1.07 (0.005)& 5.80 (0.004)& 4850 &2.77  &53.28 (6.87) &1.99 (0.31) &84.2 (3.8)  &10.4  (1.0)\\
76569 &1.06 (0.006)& 5.83 (0.004)& 4830 &2.78  &56.77 (8.03) &1.88 (0.31) &87.3 (4.7)  &10.8  (1.0)\\
77888 &1.12 (0.013)& 7.71 (0.008)& 4690 &2.63  &20.61 (3.96) &1.33 (0.27) &129.5 (10.9)&\,\,\,7.0   (0.8)\\
83224 &1.09 (0.018)& 7.34 (0.013)& 4880 &2.91  &17.89 (3.35) &1.75 (0.19) &105.7 (8.6) &\,\,\,6.1   (0.7)\\
101911&1.01 (0.007)& 6.47 (0.005)& 4885 &2.97  &16.06 (1.91) &1.63 (0.19) &74.4 (2.8)  &\,\,\,5.7   (0.5)\\
103836&1.10 (0.007)& 5.95 (0.005)& 4740 &2.89  &24.10 (2.75) &1.44 (0.32) &67.3 (2.2)  &\,\,\,7.4   (0.6)\\
104148&1.04 (0.007)& 5.70 (0.005)& 4805 &2.45  &56.70 (8.53) &1.95 (0.37) &92.4 (5.5)  &11.1  (1.1)\\
106055&1.11 (0.015)& 7.17 (0.010)& 4770 &2.68  &33.44 (7.64) &1.92 (0.27) &139.1 (14.5)&\,\,\,8.5   (1.1)\\
107122&0.96 (0.014)& 7.20 (0.010)& 4965 &3.27  &13.07 (2.28) &1.70 (0.14) &91.1 (6.7)  &\,\,\,4.9   (0.6)\\
\hline \vspace{-0.3cm} \\
\hline\hline
\end{tabular}
\end{table*}


\section{Observations and data reduction \label{sec2}}

We observed 24 giant stars that were part of the EXPRESS project. All of the targets 
are brighter than V\,=\,8 and are observable from the Southern Hemisphere. The target selection was performed according 
to their position in the HR diagram (0.8\,$\leq$\,B-V\,$\leq$\,1.2,\,-0.5\,$\leq$\,M$_V$\,$\leq$\,4.0). For more details see Jones et al. (\cite{JON11}).
\newline \indent
The data were taken using different high-resolution spectrographs, namely FEROS (Kaufer\,et\,al.\,\cite{kaufer}),
FECH, CHIRON (Tokovinin\,et\,al.\,\cite{toco}), and PUCHEROS (Vanzi\,et\,al.\,\cite{vanzi}).
In addition, we included observations taken with UCLES (Diego\,et\,al.\,\cite{diego}) as part of the Pan-Pacific Planet Search
(PPPS; Wittenmyer\,et\,al.\,\cite{WIT11}), and we complemented our data with HARPS
(Mayor\,et\,al.\,\cite{mayor2003}) archival spectra.
A brief description of these instruments is given in Table\,\ref{tb1:Instruments}.\newline\indent
For FEROS and HARPS data, the RVs were computed using the simultaneous calibration
method (Baranne\,et\,al.\,\cite{BAR96}).
For FEROS spectra, we computed the cross correlation (Tonry\,\&\,Davis\,\cite{tonry}) using a high-resolution
template of the same star (see Jones et al. \cite{jones2013}), while for HARPS spectra we used the ESO pipeline, which uses a numerical mask as template.
\newline\indent
For PUCHEROS spectra, the Doppler shift was computed in a similar way as for the FEROS, but the instrumental drift was computed from a lamp observation 
taken before and after the stellar spectrum. \newline\indent
For FECH, CHIRON, and UCLES data, the RVs were computed using the I$_{2}$ cell
method (Butler\,et\,al.\,\cite{BUT96}). The iodine cell superimposes thousands of absorption lines in the stellar light, which 
are used to obtain a highly accurate wavelength reference.
For FECH and CHIRON data, we computed the RVs following the procedure described in Jones\,et\,al.\,(\cite{jones2013}), 
while for UCLES the velocities were obtained using the Austral code (Endl\,et\,al.\,\cite{endl}), following
Wittenmyer\,et\,al.\,(\cite{WIT15}).\newline\indent
The RV precision for FEROS, CHIRON and UCLES is typically better than 5 m\,s$^{-1}$, for FECH it is 10-15 m\,s$^{-1}$, and for
PUCHEROS spectra the precision is $\sim$\,150\,m\,s$^{-1}$.

\section{Stellar properties\label{sec3}}

The main stellar properties of the primary stars are summarized in Table\,\ref{tb2:stellar}.
The visual magnitude and B-V color were taken from the Hipparcos catalog (Perryman\,et\,al.\,\cite{perry}).
A simple linear transformation was applied from the Tycho B$_{T}$ and V$_{T}$ magnitudes to B and V magnitudes in the Johnson 
photometric system, and are given by: V\,$\simeq$\,V$_{T}$\,-\,0.090\,(B$_T$\,-\,V$_T$) and
B-V\,$\simeq$\,0.850\,(B$_T$\,-\,V$_T$). The uncertainties were derived from the error in these 
transformations.
Their distances were computed using the Hipparcos parallaxes ($\Pi$). All of these objects are
relatively bright (V\,$<$\,8\,mag), and they reside at a distance  d\,$<$\,200\,pc from the Sun. \newline \indent
To derive the spectroscopic atmospheric parameters, we used the equivalent widths of a set of neutral and singly ionized iron lines.
We used the MOOG code (Sneden\,\cite{SNE73}), which solves the radiative transfer equation, using a list of atomic
transitions along with a stellar atmosphere model from Kurucz (\cite{Kur93}). 
For further details see Jones\,et\,al.\,(\cite{JON11,JON15}).\newline \indent
The stellar luminosities were computed using the bolometric correction (BC)
presented in Alonso\,et\,al.\,(\cite{ALO99}). Additionally, we corrected the visual magnitudes using the interstellar extinction maps of 
Arenou\,et\,al.\,(\cite{ARE92}). The uncertainty in the luminosity was obtained by formal propagation of the
errors in V, $\Pi$, A$_{\rm v}$ and the BC. 
The stellar mass and radius were derived by comparing the position
of these two quantities with the Salasnich\,et\,al.\,(\cite{SAL00}) evolutionary models, and their uncertainties were obtained from the 
standard deviation of 1000 random realizations, assuming Gaussian distributed errors in M$_\star$ and R$_\star$.
We adopted an uncertainty of 100 K in the effective temperature. We obtained this value by comparing our results with T$_{\rm eff}$
measurements from different studies (Jones\,et\,al.\,\cite{JON11}).
These objects cover a wide range in luminosities\,($\sim$\,5\,-\,70 L$_\odot$) and
stellar radii\,($\sim$\,3\,-\,11\,R$_\odot$), showing the wide spread
in their stellar evolutionary stages across the red giant and horizontal branch.

\subsection{Unseen companions}

\begin{figure}
   \centering
   \includegraphics[width=8cm,angle=0]{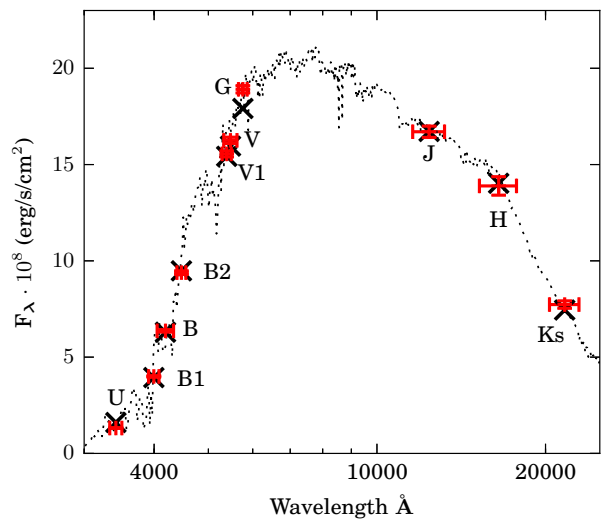}
      \caption{Example SED fit for HIP4618. The best-fitting binary model is shown with black dots,
 with the integrated model photometry in black crosses. The observed photometry is shown in red, with the horizontal error bar the width of the pass band. 
 The parameters for this fit are for the giant \,T$_{\rm eff}$=4750(100)\,K, $log{\rm g}$=2.91, radius=6.6(0.8)\,R$_\odot$
(see Table\,\ref{tb2:stellar}), and for the companion \,T$_{\rm eff}$=5500\,K, $log{\rm g}$=4.0, radius=1.2\,R$_\odot$.}
\label{SED}
\end{figure}

To determine whether features of the companion can be found in the spectrum, the contribution of the companion to the total luminosity was calculated
based on the photometric spectral energy distribution (SED). The photometry used in this procedure is Johnson, Stromgren and
2MASS photometry obtained from the literature. For each object at least five photometric measurements were found.
The SED fitting procedure used is the binary SED fit outlined in Vos\,et\,al.\,(\cite{Vos2012}) and Vos\,et.\,al.\,(\cite{Vos2013}), in which
the parameters of the giant component are kept fixed at the values determined from the spectra, and only the
parameters of the companion are varied. For this procedure, five photometric points are enough for a reliable result.\newline\indent
The observed photometry was fit with a synthetic SED integrated from Kurucz atmosphere models (Kurucz\,et\,al.\,\cite{Kurucz1979}) ranging
in effective temperature from 3000 to 7000 K, and in surface gravity from log\,$g$=2.0 dex (cgs) to 5.0 dex (cgs). The radius
of the companion was varied from R$_{\rm  comp}$ = 0.1 to 2.0 R$_{\odot}$. The SED fitting procedure uses the grid-based approach
described in Degroote\,et\,al.\,(\cite{Degroote2011}), were 1\,000\,000 models are randomly picked in the available parameter space. 
The best-fitting model is determined based on the $\chi^2$ value.\newline\indent
As the parameters (effective temperature, surface gravity, and radius) of the giant component are fixed at the values
determined from the spectroscopy and the distance to these systems is known accurately from the Hipparcos
parallax (see Table\,\ref{tb2:stellar}), the total luminosity of the giant is fixed. This allows accurately determining the
amount of missing light from the SED fit. For two systems, HIP4618 (see Fig.\,\ref{SED}) and HIP59367, the SED fit shows
that about 4-5\% of the total light originates from the companion. For all other systems the contribution of
the companion to the total light is lower than 1\%. This contribution is too low for spectral separation to work or to determine 
in any way reliable parameters for the companion star.\newline\indent
None of the model SEDs based on the spectroscopically obtained parameters shows a surplus luminosity compared to
literature photometry. This is an additional indication of the correctness of the giant companion's spectroscopic parameters.

\section{Orbital elements}
\label{sec4}

In this section we analyze the orbital properties of the 24 binary systems. We separate these systems according to their
orbital period into three groups: i) systems for which the observational time span is longer than the orbital period and for wich
thus a reliable orbital solution can be derived, 
ii) systems with longer orbital periods, for which it is possible to obtain a solution, but with a high level of degeneracy
in the orbital parameters, and iii) systems that present a RV trend. 

 
\begin{table*}
\centering
  \caption[]{Orbital elements of the binary companions. The systems with (*) a have a full orbital coverage. Mass function f(M)=$m_2^2$ sin$^3 i$/(m$_1$+m$_2)^2$}
  \label{tb3_Table_fita}
\begin{tabular}{lrrrrrr}
\hline\hline\
                             & HIP4618*       & HIP10548*      & HIP59367       & HIP83224*       & HIP73758*     & HIP104148     \\
\hline \vspace{-0.3cm} \\
$P$ (days)                   & 211.4 (0.03)   &429.1 (0.25)    &2779.3 (84.29)  & 173.3 (0.03)    &97.1 (0.002)   & 1599.3 (8.15) \\
T$_0$ (JD-2450000)           & 5406.3 (0.32)  &5306.6 (1.51)   &4832.7 (13.07)  & 5251.3 (0.1)    &5304.3 (0.03)  & 3878.5 (3.59) \\
$e$                          & 0.1 (0.003)    &0.3 (0.003)     &0.8 (0.13)      & 0.3 (0.001)     &0.4 (0.0006)   & 0.2 (0.00)    \\
$\omega$ (deg)               & 40.8 (0.58 )   &5.0 (1.86)      &235.7 (15.68)   & 85.6 (0.29)     &53.4 (0.08)    & 194.4 (2.40)   \\
$K$ (m\,s$^{-1}$)            & 12939.2 (7.84 )&6942.8 (29.82)   &5947.6 (182.71)& 8742.5 (6.91)   &11817.4 (11.00)& 5581.3 (5.42)  \\
f(M)\,(10$^{-3}$M$_{\odot}$) & 46.7 (0.09)    &12.9 (0.17)      &13.1 (6.9)     & 10.4 (0.03)     &12.8 (0.04)    & 27.1 (0.16)    \\
\hline \vspace{-0.3cm} \\
\hline\hline
\end{tabular}
\end{table*}

\subsection{Short-period binaries \label{sec4.1}}

Four of the 24 stars, show large RV variations ($\gtrsim$\,10\,km\,s$^{-1}$), with orbital periods $P$\,$\lesssim$\,430 
days. For these, it was possible to fully constrain the orbital solution. 
To determine the orbital elements of the systems, we used the 2.17 version of the Systemic Console
(Meschiari\,et\,al.\,\cite{meschiari}), excluding the PUCHEROS velocities, which have uncertainties up to $\sim$\,100 times larger
than UCLES, FEROS, and CHIRON data. The stars HIP4618, HIP10548, HIP73758, and HIP83224 have periods shorter than $\sim$\,430\,days.
The orbital elements of the four stellar companions are listed in Table\,\ref{tb3_Table_fita}. Figure \,\ref{FBinaries_sol} shows the resulting RV
curves. In the four cases, the RV data cover more than two orbital periods.

\subsection{Long-period binaries}
\label{sec:4.2}

\begin{figure}
   \centering
   \includegraphics[width=7cm,angle=-90]{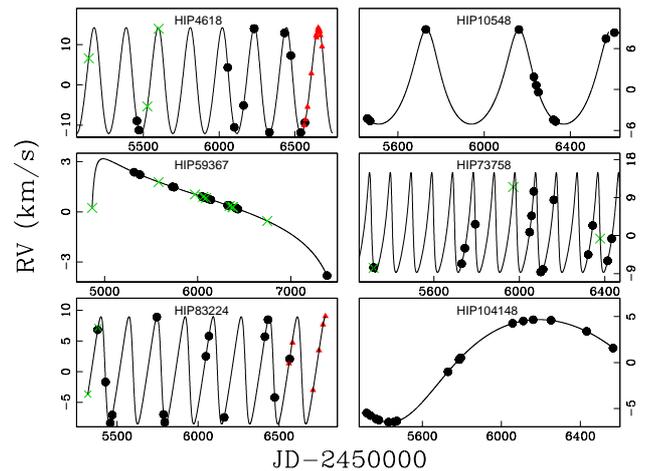}
      \caption{RV variations of the six binary systems with reliable orbital solution. 
      The black circles, blue squares, green crosses, and red triangles correspond
       to FEROS, CHIRON, UCLES, and PUCHEROS data, respectively. The best orbital solutions are overplotted (solid line).
       The post-fit RMS is typically $\sim$\,10\,m\,s$^{-1}$.}
         \label{FBinaries_sol}
   \end{figure}

\begin{figure}
   \centering
   \includegraphics[width=7cm,angle=-90]{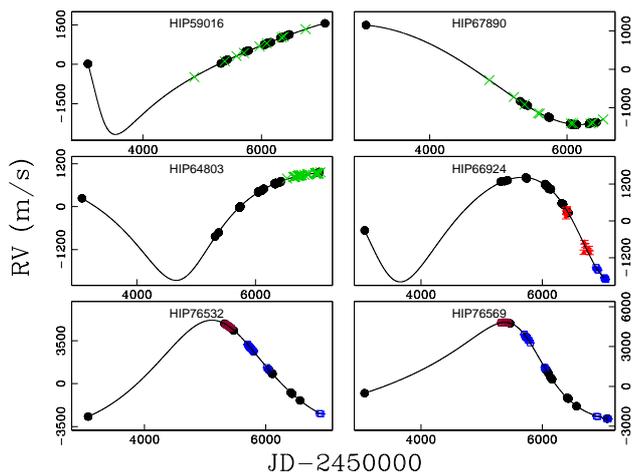}
      \caption{RV variations of six long-period binaries. The black circles, blue squares, green crosses, brown open stars, and red triangles correspond
       to FEROS, CHIRON, UCLES, FECH, and PUCHEROS data, respectively. In each case, one possible orbital solution is overplotted (solid line).}
         \label{FBinaries_ds_rv}
   \end{figure}

In eight cases, we observe large RV variations, but with orbital periods exceeding the observational time span.
However, for  HIP\,59367 and HIP\,104148 the phase coverage is good enough to obtain a unique orbital 
solution. Figure\,\ref{FBinaries_sol} shows the RV curves of these two stars.
The orbital elements of the binary companions are listed in Table\,\ref{tb3_Table_fita}.\newline
For the remaining six cases the orbital solution is partially degenerated 
because of the poor phase coverage, meaning that we can only set lower and upper limits for the orbital period and the eccentricity.
Figure\,\ref{FBinaries_ds_rv} shows the RV curve of these stars. 
One possible orbital solution is overplotted.  In these cases it is not possible to unambiguously obtain a solution.

\subsubsection{Long-period trends}
\label{sec4.3}

\begin{figure}
\centering
\includegraphics[width=7cm,angle=-90]{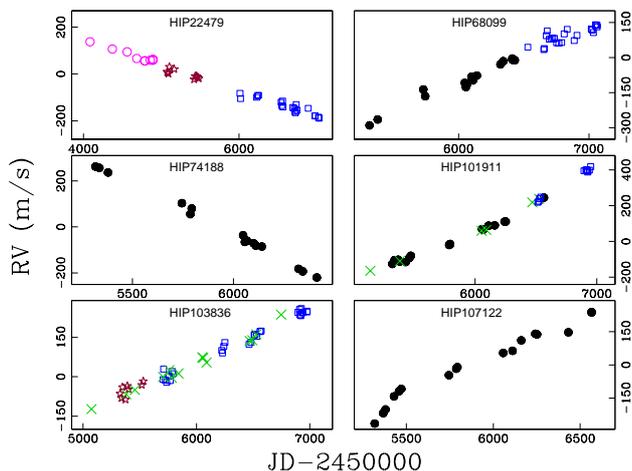}
\caption{RV variations of six long-period brown dwarf companion candidates. 
The black filled circles, blue squares, green crosses, brown open stars, and magenta open circles correspond to FEROS, CHIRON, UCLES, FECH,
 and HARPS data, respectively.
         } \label{FBDC_lt}
   \end{figure}

\begin{figure}
\centering
\includegraphics[width=7cm,angle=-90]{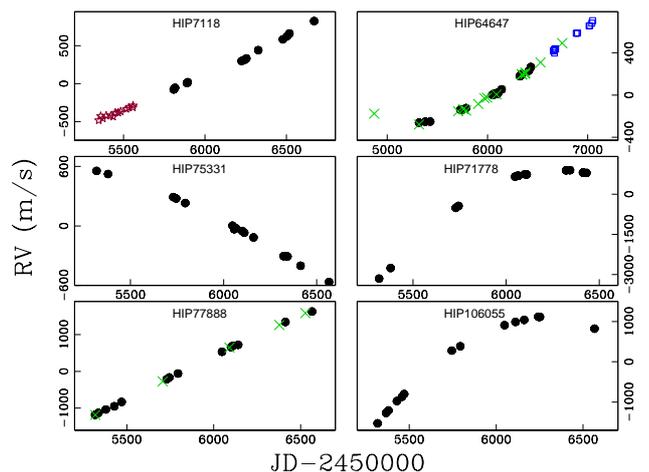}
\caption{Long-period RV trends of six binary systems. The black circles, blue squares, brown open stars, and green
crosses correspond to FEROS, CHIRON, FECH, and UCLES data, respectively.} \label{FBinaries_rvtrends}
   \end{figure}

The remaining 12 stars in this sample present RV variations, ranging from thousands of m\,s$^{-1}$ level up to
peak-to-peak variations of several km\,s$^{-1}$. Half of the systems present a linear RV trend, while the remainder show
some level of curvature in the observed velocities. 
Figure\,\ref{FBDC_lt} shows the RV epochs of the six stars that present the smallest RV variations ($\sim$\,500\,m\,s$^{-1}$). 
Since these stars show moderate RV variations, they are candidates for hosting long-period brown dwarfs, 
which makes them very interesting targets for direct imaging to determine the nature of the companion. 
Finally, Fig.\,\ref{FBinaries_rvtrends} 
shows the velocity variations of the stars that present large RV long-trend variations ($\gtrsim$\,1\,k\,m\,s$^{-1}$), 
which are most likely part of a long-period stellar binary system.

\begin{figure}
\centering
\includegraphics[width=7cm,angle=-90]{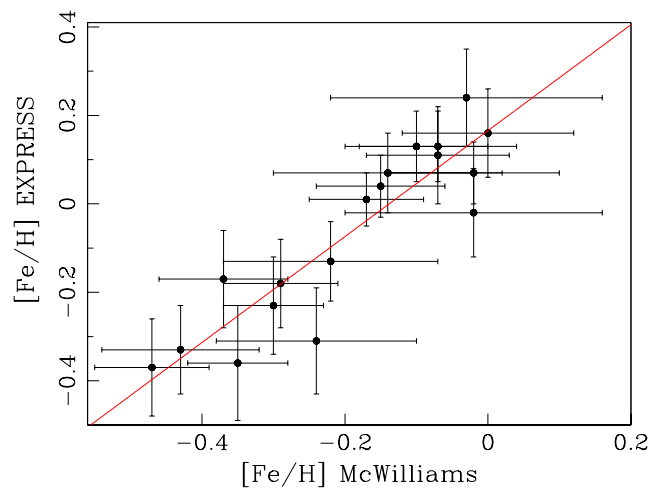}
\caption{EXPRESS versus MCW90 metallicities for the 18 targets in common. The red line corresponds to the best linear fit. \label{Ex_Mc}}
\end{figure}

\section{Metallicity distribution \label{sec5.1}}

\begin{figure}
\centering
\includegraphics[width=7cm,angle=-90]{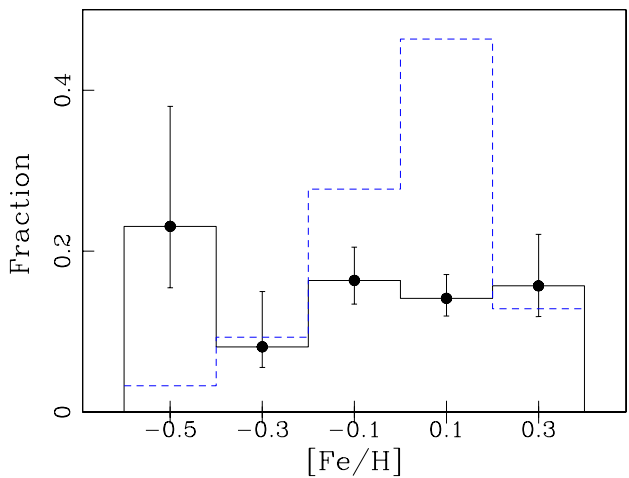}
\caption{Normalized histogram of the metallicity distribution of the primary stars (black solid line). The overall metallicity
distribution of the parent sample is overplotted (dashed blue line). The width of the bins is 0.2 dex. 
} \label{Fmetallicity}
\end{figure}

\begin{figure}
\centering
\includegraphics[width=7cm,angle=-90]{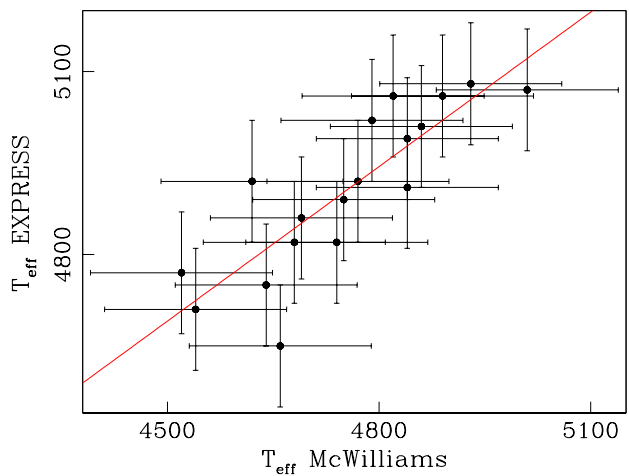}
\caption{EXPRESS versus MCW90 effective temperatures (T$_{\rm eff}$) for the 18 targets in common. The red line corresponds to the best linear fit.}
 \label{Teff}
\end{figure}

We investigated the metallicity distribution of the primary giant stars and their binary
fraction. Additionally, we included 82 stars from Setiawan et al. (\cite{setiawan}) and 150 stars from MAS08,
wich makes up a sample of 395 giant stars in this analysis.\newline\indent
We computed the metallicities of SET04 targets using FEROS\,archival data.
For the MAS08 targets, we used only those targets with metallicities computed by McWilliam\,(\cite{mcwilliam}; MCW90 hereafter). 
We compared our sample with the MCW90 sample and we found 18 common stars. These stars are shown in Fig.\,\ref{Ex_Mc}.
To remove any bias due to differences in the metallicity derived by our method and MCW90, we adjusted a linear function to correlate the two studies. 
We found a linear 
correlation of the form [Fe/H]$_{EXP}$\,=\,1.20\,[Fe/H]$_{MCW90}$+0.17. Figure \ref{Ex_Mc} shows the EXPRESS versus MCW90 metallicities for the 18 
targets in common. The best linear fit is overplotted. The RMS of the fit is 0.08 dex, and the Pearson linear coefficient is $r$=0.90.\newline\indent
Using this information, we converted from MCW90 metallicities into our metallicity scale, for all of the binaries listed in MAS08 and metallicities from
MCW90. Figure\,\ref{Fmetallicity} shows the normalized metallicity distribution of the primary stars for a total of 59 binaries, including EXPRESS, 
SET04, and MAS08 systems (black solid line). The error bars were computed according to Cameron (\cite{CAM11}).
The overall sample distribution is overplotted (blue dashed line).
The metallicity distribution between $\sim$ -0.3 to 0.3 dex is nearly flat. Moreover, the highest fraction is 
obtained at metallicities around -0.5 dex. 
Interestingly, Raghavan et al.\,(\cite{RAG10}) showed that binary systems among solar-type stars redder than B-V = 0.625 are 
more frequent around stars with [Fe/H] $\lesssim$ -0.3 dex, in good agreement with our findings\footnote{Most of the stars in the combined sample have B-V > 0.8}.
However, we note that the bin centered on -0.5 dex in the one with the least
number of stars in the combined sample  (13 systems), and therefore with the largest error bars.
This result is in stark contrast with the observed metallicity distribution of planet-hosting giant stars, showing a strong increase in the giant planet 
frequency with increasing metallicity (e.g., Reffert et al.\,\cite{reffert}; Jones et al. \cite{jones16}), as found in solar-type stars 
(e.g., Fischer\,\&\,Valenti\,\cite{FIS05}). This observational result also agrees with recent hydrodynamical simulations
showing that the formation of binary systems across a wide range of stellar masses is not significantly 
affected by the metal content of the molecular clouds (Bate \cite{BAT14}).\newline\indent
Additionally, we also studied the relation in the effective temperatures (T$_{\rm eff}$) for these 18 stars in common. 
We compared the McWilliam and our T$_{\rm eff}$ values and found a linear correlation of the form T$_{\rm eff}$(EXP)\,=\,0.84\,T$_{\rm eff}$(MCW90)+889.1.
The Pearson linear coefficient is $r$=0.84. Figure\,\ref{Teff} shows the MVW90 versus EXPRESS T$_{\rm effs}$.
The linear fit is overplotted. The error bars are not given in the MC90, and we used error bars given by the
standard deviation for de T$_{\rm eff}$(MC90) values.
Our values overestimate those derived by MCW90, which mainly explains why we also see a systematic
difference in the derived metallicities (see Fig.\,\ref{Ex_Mc}).

\section{Period-eccentricity distribution \label{sec5.2}}

  \begin{figure}
   \includegraphics[width=7cm,angle=-90]{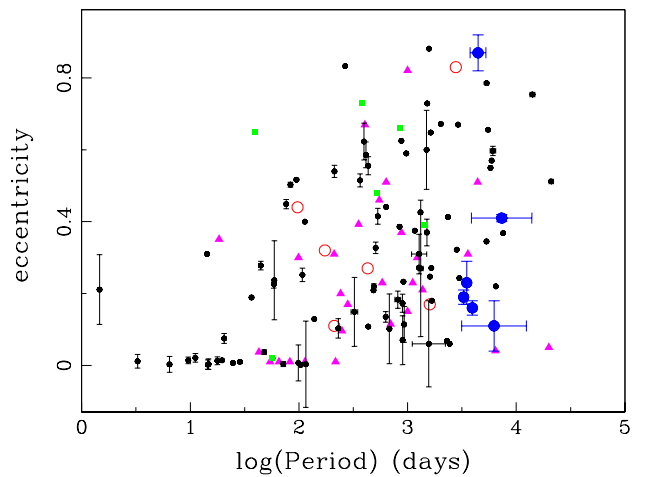}
      \caption{Orbital period versus eccentricity for 130 spectroscopic giant binary stars. The red open circles represent our binaries with 
               well-constrained orbital periods, while the blue filled circles represent those with poorly constrained orbits.
               The black dots, magenta filled triangles, and green filled squares correspond to MAS08, VER95, and SET04 binaries, respectively.}
         \label{p_e}
   \end{figure}

We studied the period-eccentricity distribution of 130 spectroscopic binary systems in giant stars. We included all of the binaries 
with orbital solution from VER95, SET04, MAS08, and those presented here for which it was possible to obtain an orbital solution. 
Figure\,\ref{p_e} shows the orbital period versus eccentricity of these systems. Systems with short orbital 
periods (P\,$\lesssim$\,20\,days) present nearly circular orbits, similar to solar-type binaries, which is most likely explained by tidal circularization. 
Furthermore, there is a transition region from P\,$\sim$\,20-80\,days where 
binary systems present moderate eccentricities ($e\,\lesssim$\,0.4). Finally, at longer orbital periods, there is a wide spread in
$e$, ranging from nearly circular orbits to highly eccentric systems ($e \sim$ 0.9). This transition region and eccentricity distribution at 
orbital periods longer than $\sim$\,100\,d is also observed in solar-type binaries (e.g., Duquennoy \& Mayor \cite{DUQ91}; 
Jenkins et al. \cite{JEN15}).\newline\indent 
The giant binary systems with moderately long orbital periods ($\sim$400-800\,days) might be the precursors 
of the wide eccentric hot subdwarf binaries studied by Vos\,et\,al.\,(\cite{Vos2015}).
A hot subdwarf star is a core helium-burning star located at the blue end of the horizontal branch, with colours 
similar to main-sequence B stars, but with much broader Balmer lines (Sargent\,\&\,Searle\,\cite{Sar68}).
The study of the orbital parameters of these systems will therefore be useful in binary population synthesis studies for wide sdB binaries
to help determine the correct evolutionary channels of these evolved binaries.


\section{Summary \label{sec6}}

We presented a sample of 24 giant stars that have revealed large radial velocity
variations, which are induced by massive substellar or stellar companions. Based on precision RVs computed from high-resolution spectroscopic 
observations obtained as part of the EXPRESS program, we were able to fully constrain the orbital elements of 6 systems. 
In 6 more cases, we were able to obtain a solution although the orbital elements are poorly constrained. 
In the remaining cases, the primary star exhibits a RV trend, thus no solution is obtained. Six of these  stars present velocity variations that 
are compatible with the presence of a long-period brown dwarf companion.  \newline \indent
In addition, we studied the metallicity distribution of the primary stars. For this purpose, 
we retrieved literature data from two different studies, namely MAS08 and SET04. Our final sample comprised 59 spectroscopic binary systems,
detected from a parent sample of 395 giant stars, covering a range of metallicities between [Fe/H] $\sim$ 
-0.5 and +0.5 dex. We found no significant correlation between the frequency of binary companions and the stellar metallicity. 
This result reinforces the fact that stellar binaries are formed mainly by gravitational collapse, which is highly insensitive to the
 dust content of the protostellar disk (e.g., Bate \cite{BAT14}), while planetary systems, including those orbiting giant stars, are
 formed in the protoplanetary disk by the 
core-accretion mechanism (e.g., Gonzalez, \cite{GON97}; Santos et al. \cite{SAN01}; Reffert et al. \cite{reffert}; Jones et al. \cite{jones16}).
\newline \indent 
Finally, we studied the period-eccentricity distribution of the companions. We included a total of 130 spectroscopic binaries from the literature 
with known eccentricities. We found an eccentricity distribution that is characterized by short-period systems (P$\lesssim$\,20\,days) that present 
very low eccentricities (with the exception of one case with $e$\,$\sim$\,0.2). For orbital periods between $\sim$\,20-80 days, all of the systems
present moderate orbital eccentricities ($e\,\lesssim$\,0.4). At longer orbital periods, there is a wide spread in $e$, from nearly circular orbits 
to eccentricities as high as $\sim$\,0.9. The overall distribution is qualitatively similar to the distribution observed in solar-type stars, although
the circularization edge is found at slightly longer orbital periods, which is most likely explained by the stronger tidal effect induced by the larger 
stellar radii.

\begin{acknowledgements}

MJ acknowledges support from FONDECYT project \#3140607 and FONDEF project \#CA13I10203\\
L.Vanzi acknowledges PUCHEROS was funded by CONICYT through projects FONDECYT n. 1095187 and n. 1130849.\\
H.Drass acknowledges financial support from FONDECYT project \#3150314\\
R.E.M. acknowledges support by VRID-Enlace 214.016.001-1.0 and the BASAL Centro de 
Astrof{\'{i}}sica y Tecnolog{\'{i}}as Afines (CATA) PFB--06/2007.\\
FOE acknowledges support from FONDECYT through postdoctoral grant 3140326 and from
project IC120009 ``Millennium Institute of Astrophysics (MAS)'' of the Iniciativa
Cient\'ifica Milenio del Ministerio de Econom\'ia, Fomento y Turismo de Chile.

\end{acknowledgements}


\begin{appendix}
\section{Radial velocity tables}

\begin{table}[H]
\centering
  \caption{Radial velocity variations for HIP4618.}
  \label{tb1:instruments}
\begin{tabular}{rrrrr}
\hline\hline
JD      & RV         & err       & Instrument \\ 
-2450000& (ms$^{-1}$)  & (ms$^{-1}$) &      \\
\hline \vspace{-0.3cm} \\
5140.11 & -3.4 & 1.8 & UCLES\\
5525.95 & -11966.3 & 2.7 & UCLES\\
5601.93 & 7339.9 & 3.3 & UCLES\\
5457.74 & -7583.3 & 5.2& FEROS\\
5470.78 & -9871.8 & 7.3 & FEROS\\
6056.93 & 5693.4 & 5.5 & FEROS \\
6099.92 & -9187.2 & 6.3 & FEROS\\
6160.95 & -3752.3 & 7.7 & FEROS\\
6230.65 & 15340.8 & 5.0  &FEROS\\
6231.79 & 15371.9 & 4.5 & FEROS\\
6331.53 & -10427.0 & 7.7 & FEROS\\
6431.93 & 14269.6 & 6.3 & FEROS\\
6472.95 & 8653.4 & 5.5 & FEROS\\
6537.92 & -10493.3 & 4.3 & FEROS\\
6565.72 & -8014.1 & 5.4 & FEROS\\
6561.57 & -6387.6 & 121.2 & PUCHEROS\\
6569.73 & -4962.8 & 106.7 & PUCHEROS\\
6582.71 & -1780.8 & 76.0 & PUCHEROS\\
6608.63 & 6535.8 & 90.2 & PUCHEROS\\
6638.57 & 15857.5 & 131.8 & PUCHEROS\\
6643.63 & 16447.2 & 199.5 & PUCHEROS\\
6650.58 & 17404.8 & 144.9 & PUCHEROS\\
6653.59 & 17881.8 & 136.9 & PUCHEROS\\
6660.56 & 17340.9 & 115.1 & PUCHEROS\\
6664.58 & 16870.5 & 145.9 & PUCHEROS\\
6668.61 & 15979.9 & 124.2 & PUCHEROS\\
6678.55 & 13211.7 & 110.6 & PUCHEROS\\
\hline \vspace{-0.3cm} \\
\hline\hline
\end{tabular}
\end{table}

\begin{table}[]
\centering
  \caption{Radial velocity variations for HIP7118.}
  \label{tb1:instruments}
\begin{tabular}{rrrrr}
\hline\hline
JD      & RV         & err       & Instrument \\
-2450000& (ms$^{-1}$)  & (ms$^{-1}$) &      \\
\hline \vspace{-0.3cm} \\
  5347.90&   -484.3&   14.3&FECH\\ 
  5359.89&   -428.9&   14.4&FECH\\ 
  5373.86&   -460.1&   13.0&FECH \\ 
  5393.85&   -422.5&   16.5&FECH \\ 
  5421.75&   -422.7&   13.5&FECH \\ 
  5435.71&   -436.0&   14.0&FECH \\ 
  5449.67&   -385.1&   15.8&FECH \\ 
  5467.64&   -382.8&   12.6&FECH \\ 
  5482.64&   -366.9&   21.4&FECH \\ 
  5517.62&   -336.0&   13.3&FECH \\ 
  5531.62&   -316.0&   13.3&FECH \\ 
  5557.61&   -314.3&   14.3&FECH \\ 
  5558.62&   -287.6&   15.0&FECH \\ 
  5807.93&    -76.4&    7.7&CHIRON \\ 
  5815.88&    -55.1&   10.6&CHIRON \\ 
  5888.74&      7.3&    8.0&CHIRON \\ 
  5893.62&     19.3&    7.9&CHIRON \\ 
  6224.60&    299.0&    8.1&CHIRON \\ 
  6244.61&    316.6&    6.8&CHIRON \\ 
  6253.64&    332.5&    7.5&CHIRON \\ 
  6326.56&    441.4&    7.2&CHIRON \\ 
  6477.89&    585.3&    6.9&CHIRON \\ 
  6505.91&    631.1&    7.2&CHIRON \\ 
  6517.85&    663.2&    8.0&CHIRON \\ 
  6669.57&    825.6&    6.8&CHIRON \\ 
  6870.80&   1053.2&    7.0&CHIRON \\
\hline\hline
\end{tabular}
\end{table}

\begin{table}[]
\centering
  \caption{Radial velocity 10548.}
  \label{tb1:instruments}
\begin{tabular}{rrrrr}
\hline\hline
JD      & RV         & err       & Instrument \\
-2450000& (ms$^{-1}$)  & (ms$^{-1}$) &      \\
\hline \vspace{-0.3cm} \\
5457.79&  -5857.7&   5.2& FEROS\\
5470.81&  -6206.9&   6.3&FEROS\\ 
5729.92&   7223.3&  10.4&FEROS\\ 
6160.86&   7205.0&   2.5&FEROS\\ 
6230.71&    262.4&   5.5&FEROS\\ 
6241.73&   -963.4&   5.3&FEROS\\ 
6251.75&  -1975.6&   4.7&FEROS\\ 
6321.57&  -6032.3&   7.1& FEROS\\ 
6331.58&  -6282.4&   7.4&FEROS\\ 
6565.75&   5874.6&   5.7&FEROS\\ 
6604.67&   6752.9&   4.7&FEROS \\
\hline\hline
\end{tabular}
\end{table}

\begin{table}[]
\centering
  \caption{Radial velocity variations for HIP22479.}
  \label{tb1:instruments}
\begin{tabular}{rrrrr}
\hline\hline
JD      & RV         & err       & Instrument \\
-2450000& (ms$^{-1}$)  & (ms$^{-1}$) &      \\
\hline \vspace{-0.3cm} \\
4085.64 & 107.4& 0.2&HARPS\\
4371.79 & 76.7 & 0.3&HARPS\\
4561.49 & 64.6 & 0.2&HARPS \\
4684.91 & 36.3 & 0.3&HARPS \\
4787.76 & 26.4 & 0.3&HARPS \\
4788.79 & 25.6 & 0.2&HARPS \\
4858.55 & 29.5 & 0.3&HARPS \\
4891.58 & 28.4 & 0.3&HARPS \\
4893.59 & 33.2 & 0.4&HARPS \\
5075.87 & 102.1& 12.3&FECH\\
5084.91 & 97.8 & 11.9& FECH \\
5093.92 & 94.2 & 11.7& FECH \\
5104.82 & 122.8& 12.5& FECH \\
5166.71 & 112.6& 10.8& FECH \\
5421.89 & 69.2 & 12.6& FECH \\
5435.87 & 82.4 & 11.7& FECH \\
5449.89 & 83.6 & 11.9& FECH \\
5467.79 & 79.3 & 11.2& FECH \\
5482.74 & 73.2 & 14.4& FECH \\
6011.55 & 9.5  & 9.2& CHIRON \\
6019.52 & -13.0& 6.7& CHIRON\\
6226.82 & -7.0 & 6.3& CHIRON \\
6239.72 & -0.7 & 5.6& CHIRON \\
6248.69 & 1.3 & 4.8& CHIRON \\
6539.88 & -29.0 & 9.7& CHIRON\\
6547.89 & -42.8 & 6.3& CHIRON \\
6557.82 & -23.9 & 6.3& CHIRON \\
6563.72 & -48.9 & 6.1& CHIRON \\
6690.58 & -51.0 & 5.0& CHIRON \\
6711.54 & -54.3 & 5.6& CHIRON \\
6720.62 & -67.6 & 8.8& CHIRON \\
6723.50 & -72.8 & 5.6& CHIRON \\
6734.49 & -38.9 & 5.6& CHIRON \\
6745.48 & -62.8 & 5.7& CHIRON \\
6888.92 & -54.2 & 6.1& CHIRON \\
6976.79 & -86.2 & 5.2& CHIRON \\
6976.79 & -86.4 & 5.4& CHIRON \\
7019.69 & -92.4 & 5.8&  CHIRON \\
7027.66 & -96.3 & 5.6& CHIRON \\
\hline\hline 
\end{tabular} 
\end{table}

\begin{table}[]
\centering
  \caption{Radial velocity variations for HIP59016.}
  \label{tb1:instruments}
\begin{tabular}{rrrrr}
\hline\hline
JD      & RV         & err       & Instrument \\
-2450000& (ms$^{-1}$)  & (ms$^{-1}$) &      \\
\hline \vspace{-0.3cm} \\
3072.75 & -723.8 & 6.6&FEROS\\
5317.51 & -707.8 & 5.1&FEROS\\
5379.50 & -617.9 & 3.1&FEROS\\
5428.48 & -562.7 & 5.1& FEROS\\
5729.54 & -255.8 & 10.2&FEROS \\
5744.49 & -244.5 & 5.6&FEROS \\
5786.48 & -216.4 & 4.5&FEROS\\
6047.52 & 11.4 & 5.6&FEROS \\
6056.51 & 1.3 & 5.5&FEROS \\
6099.50 & 69.3 & 4.8&FEROS \\
6110.48 & 56.6 & 4.0&FEROS \\
6140.52 & 80.1 & 4.6&FEROS \\
6160.47 & 98.5 & 4.2&FEROS \\
6321.69 & 262.7 & 7.6&FEROS \\
6331.68 & 247.6 & 7.8&FEROS \\
6342.64 & 288.5 & 7.8&FEROS \\
6412.53 & 322.3 & 4.3&FEROS \\
6412.73 & 323.2 & 4.7&FEROS \\
6431.57 & 355.8 & 5.7&FEROS \\
6472.54 & 394.8 & 4.9&FEROS \\
7072.78 & 816.7 & 8.7&FEROS \\
4866.23 & -1166.6 & 2.5&UCLES\\
5380.88 & -563.1 & 2.4&UCLES\\
5580.23 & -367.3 & 2.8&UCLES \\
5706.91 & -244.9 & 4.8&UCLES \\
5970.19 & 0.0 & 2.9&UCLES \\
6090.97 & 119.4 & 4.3&UCLES \\
6345.14 & 339.6 & 2.9&UCLES \\
6376.09 & 374.8 & 2.6&UCLES \\
6747.01 & 659.5 & 2.3&UCLES\\
\hline\hline
\end{tabular}
\end{table}

\begin{table}[]
\centering
  \caption{Radial velocity variations for HIP59367.}
  \label{tb1:instruments}
\begin{tabular}{rrrrr}
\hline\hline
JD      & RV         & err       & Instrument \\
-2450000& (ms$^{-1}$)  & (ms$^{-1}$) &      \\
\hline \vspace{-0.3cm} \\
4866.22 & -112.0 & 2.5&UCLES\\
5581.19 & 1436.2 & 2.4&UCLES\\
5970.19 & 698.5 & 2.8&UCLES\\
6059.98 & 542.1 & 4.8&UCLES\\
6088.92 & 486.8 & 2.9&UCLES\\
6344.14 & 0.0 & 4.3&UCLES\\
6376.04 & -62.6 & 2.9&UCLES\\
6378.01 & -66.8 & 2.6&UCLES\\
6748.08 & -897.2 & 2.3&UCLES\\
5317.53 & 1755.5 & 5.1&FEROS\\
5379.51 & 1608.1 & 3.8&FEROS\\
5729.54 & 879.4 & 6.0&FEROS\\
5744.50 & 860.4 & 5.9&FEROS\\
6047.52 & 296.8 & 3.8&FEROS\\
6056.52 & 275.2 & 3.9&FEROS\\
6066.53 & 252.2 & 5.0&FEROS\\
6099.52 & 200.2 & 5.0&FEROS\\
6110.48 & 168.0 & 4.7&FEROS\\
6140.53 & 101.7 & 6.9&FEROS\\
6321.69 & -232.5 & 4.8&FEROS\\
6331.71 & -247.6 & 4.7&FEROS\\
6342.65 & -245.0 & 4.6&FEROS\\
6412.54 & -399.5 & 4.7&FEROS\\
6412.74 & -392.8 & 5.1&FEROS\\
6431.57 & -453.1 & 4.4&FEROS\\
7388.85 & -4427.1 & 6.0&FEROS\\
\hline \vspace{-0.3cm} \\
\hline\hline
\end{tabular}
\end{table}

\begin{table}[]
\centering
  \caption{Radial velocity variations for HIP64647.}
  \label{tb1:instruments}
\begin{tabular}{rrrrr}
\hline\hline
JD      & RV         & err       & Instrument \\
-2450000& (ms$^{-1}$)  & (ms$^{-1}$) &      \\
\hline \vspace{-0.3cm} \\
4870.21 & -151.9 & 2.8&UCLES\\
5317.98 & -254.3 & 1.6&UCLES\\ 
5706.96 & -129.2 & 2.2&UCLES\\ 
5757.89 & -110.0 & 5.9&UCLES\\ 
5787.87 & -123.0 & 3.5&UCLES\\ 
5908.24 & -60.6 & 2.2&UCLES\\ 
5969.23 & -6.6 & 2.0&UCLES\\ 
5995.19 & 0.0 & 2.1&UCLES\\ 
6088.95 & 29.0 & 2.6&UCLES\\ 
6344.19 & 218.1 & 3.3&UCLES\\ 
6345.12 & 220.1 & 3.1&UCLES\\ 
6376.11 & 223.4 & 2.4&UCLES\\ 
6377.06 & 237.5 & 3.1&UCLES\\ 
6528.85 & 331.9 & 3.7&UCLES\\ 
6745.09 & 515.9 & 2.3& UCLES\\
5317.57 & -261.1 & 6.6&FEROS\\ 
5379.67 & -252.3 & 4.9&FEROS\\ 
5428.49 & -251.4 & 6.6&FEROS\\ 
5729.59 & -137.8 & 12.3&FEROS\\ 
5744.55 & -142.7 & 6.7&FEROS\\ 
5786.55 & -123.4 & 5.3&FEROS\\ 
6047.58 & 2.7 & 7.7&FEROS\\ 
6056.56 & 1.7 & 5.7&FEROS\\ 
6066.58 & 11.5 & 6.9&FEROS\\ 
6099.56 & 11.5 & 7.1&FEROS\\ 
6110.54 & 28.2 & 6.5&FEROS\\ 
6140.58 & 54.0 & 4.6&FEROS\\ 
6321.75 & 179.2 & 7.0&FEROS\\ 
6331.77 & 184.4 & 7.5&FEROS\\ 
6342.72 & 198.7 & 7.9&FEROS\\ 
6412.70 & 231.5 & 4.6&FEROS\\ 
6431.62 & 265.3 & 6.8& FEROS\\
6655.79 & -119.5 & 5.8&CHIRON\\
6660.81 & -121.9 & 6.5& CHIRON\\
6664.79 & -107.1 & 6.1& CHIRON\\
6668.85 & -141.1 & 6.5&CHIRON \\
6672.88 & -98.8 & 6.1&CHIRON\\ 
6678.80 & -103.5 & 5.6&CHIRON\\ 
6883.51 & 47.0 & 6.2&CHIRON\\ 
6894.48 & 47.8 & 6.3&CHIRON\\ 
6898.47 & 50.6 & 6.0&CHIRON\\ 
7013.85 & 116.6 & 5.3&CHIRON\\ 
7021.85 & 119.8 & 5.1&CHIRON\\ 
7039.84 & 143.3 & 4.8&CHIRON\\ 
7048.82 & 166.9 & 5.0&CHIRON\\
\hline \vspace{-0.3cm} \\
\hline\hline
\end{tabular}
\end{table}

\begin{table}[]
\centering
  \caption{Radial velocity variations for HIP64803.}
  \label{tb1:instruments}
\begin{tabular}{rrrrr}
\hline\hline
JD      & RV         & err       & Instrument \\
-2450000& (ms$^{-1}$)  &(ms$^{-1}$) &      \\
\hline \vspace{-0.3cm} \\
3072.81&   -226.8&  25.0&FEROS\\ 
5317.57&  -1296.4&   5.0&FEROS\\ 
5379.64&  -1183.7&   4.3&FEROS\\ 
5729.60&   -484.8&  12.7&FEROS\\ 
5729.60&   -485.7&  11.6&FEROS\\ 
5729.61&   -491.9&  12.5&FEROS\\ 
5744.55&   -449.3&   5.5&FEROS\\ 
5744.55&   -466.6&   5.6&FEROS\\ 
6047.58&    -60.6&   6.7&FEROS\\ 
6047.58&    -56.0&   6.3&FEROS\\ 
6047.58&    -62.2&   7.2&FEROS\\ 
6056.57&    -39.2&   5.1&FEROS\\ 
6056.57&    -38.1&   4.6&FEROS\\ 
6066.58&    -28.1&   5.3&FEROS\\ 
6066.58&    -33.2&   5.1&FEROS\\ 
6099.56&    -10.5&   5.5 &FEROS\\ 
6110.54&     -8.8&   4.5 &FEROS\\ 
6140.58&     35.3&   4.8&FEROS\\ 
6140.58&     29.9&   4.3&FEROS\\ 
6140.58&     35.3&   4.5&FEROS\\ 
6321.76&    181.3&   6.9&FEROS\\ 
6321.76&    180.6&   7.0&FEROS\\ 
6321.76&    178.8&   6.9&FEROS\\ 
6331.79&    172.2&   7.8&FEROS\\ 
6331.79&    170.1&   7.8&FEROS\\ 
6331.79&    169.0&   8.0&FEROS\\ 
6342.73&    197.2&   8.3&FEROS\\ 
6342.73&    196.4&   8.5&FEROS\\ 
6342.73&    201.8&   7.9&FEROS\\ 
6412.69&    223.8&   4.6&FEROS\\ 
6412.71&    221.8&   4.6&FEROS\\ 
6431.63&    252.5&   6.7&FEROS\\ 
7072.88&    478.6&   8.1&FEROS\\ 
7072.88&    490.4&   9.5&FEROS\\ 
7072.88&    497.4&  14.7&FEROS\\ 
7072.89&    501.3&  10.9&FEROS\\ 
7072.89&    499.6&  11.6&FEROS\\ 
7072.90&    508.7&  11.3&FEROS\\
6533.47&   -106.2&    4.9 &CHIRON\\ 
6644.85&    -65.8&   11.9 &CHIRON\\ 
6645.85&    -68.7&    5.2 &CHIRON\\ 
6648.85&    -62.7&    9.7 &CHIRON\\ 
6655.81&    -64.5&    4.3 &CHIRON\\ 
6666.84&    -57.2&    4.3 &CHIRON\\ 
6673.85&    -57.0&    4.3 &CHIRON\\ 
6707.84&    -40.5&    4.5 &CHIRON\\ 
6721.75&    -28.7&    4.3 &CHIRON\\ 
6735.61&    -24.7&    4.4 &CHIRON\\ 
6751.76&    -20.6&    4.9 &CHIRON\\ 
6769.53&    -23.1&    4.8 &CHIRON\\ 
6784.61&    -16.8&    4.6 &CHIRON\\ 
6827.52&     -2.1&    4.2 &CHIRON\\ 
6876.46&     19.8&    9.3 &CHIRON\\ 
6882.48&     11.3&    4.1 &CHIRON\\ 
6887.47&      4.4&    6.0 &CHIRON\\ 
6894.49&     14.6&    4.5 &CHIRON\\ 
7008.86&     44.7&    3.9 &CHIRON\\ 
7012.86&     33.5&    4.1 &CHIRON\\ 
7012.87&     30.9&    4.6 &CHIRON\\ 
7017.86&     28.2&    4.0 &CHIRON\\ 
\hline \vspace{-0.3cm} \\
\hline\hline
\end{tabular}
\end{table}

\begin{table}[]
\centering
  \caption{Continued table for HIP64803.}
  \label{tb1:instruments}
\begin{tabular}{rrrrr}
\hline\hline
JD      & RV         & err       & Instrument \\
-2450000& (ms$^{-1}$)  & (ms$^{-1}$) &      \\
\hline \vspace{-0.3cm} \\
7025.83&     44.8&    3.6 &CHIRON\\ 
7031.84&     49.2&    3.5 &CHIRON\\ 
7040.83&     34.5&    3.5 &CHIRON\\ 
7042.75&     41.4&    3.8 &CHIRON\\ 
7044.80&     48.6&    3.6 &CHIRON\\ 
7046.77&     45.4&    3.9 &CHIRON\\ 
7048.83&     52.7&    3.8 &CHIRON\\ 
7061.73&     27.5&    3.6 &CHIRON\\ 
7063.76&     41.5&    4.4 &CHIRON\\
7097.87&    65.7 &   3.7 &CHIRON\\ 
\hline \vspace{-0.3cm} \\
\hline\hline
\end{tabular}
\end{table}

\begin{table}[]
\centering
  \caption{Radial velocity variations for HIP66924.}
  \label{tb1:instruments}
\begin{tabular}{rrrrr}
\hline\hline
JD      & RV         & err       & Instrument \\
-2450000& (ms$^{-1}$)  & (ms$^{-1}$) &      \\
\hline \vspace{-0.3cm} \\
3072.83 & -1277.9 & 20.0&FEROS\\ 
5317.623 & 312.2 & 5.1&FEROS\\ 
5379.643 & 328.8 & 4.2&FEROS\\ 
5428.523 & 359.5 & 6.0&FEROS\\ 
5729.623 & 444.8 & 11.6&FEROS\\ 
5729.623 & 442.6 & 10.1&FEROS\\ 
5744.593 & 420.5 & 7.9&FEROS\\ 
5744.593 & 422.7 & 7.8&FEROS\\ 
6047.613 & 233.7 & 6.4&FEROS\\ 
6047.613 & 230.9 & 6.7&FEROS\\ 
6056.603 & 192.1 & 5.2&FEROS\\ 
6056.603 & 191.5 & 5.1&FEROS\\ 
6066.613 & 175.8 & 5.4&FEROS\\ 
6066.62 & 178.9 & 5.4&FEROS\\ 
6099.59 & 133.2 & 5.5&FEROS\\ 
6110.58 & 73.2 & 5.2&FEROS\\ 
6140.61 & 73.8 & 4.2&FEROS\\ 
6140.62 & 65.3 & 4.7&FEROS\\ 
6321.79 & -395.5 & 7.4&FEROS\\
6321.79 & -392.5 & 6.9&FEROS\\ 
6342.75 & -421.7 & 7.1&FEROS\\ 
6342.75 & -427.4 & 7.5&FEROS\\ 
6412.61 & -655.7 & 3.7&FEROS\\ 
6431.65 & -708.8 & 6.4&FEROS\\ 
6391.60 & -712.5 & 80.9&PUCHEROS\\ 
6398.58 & -477.5 & 61.6&PUCHEROS\\ 
6400.56 & -562.8 & 65.3&PUCHEROS\\ 
6407.59 & -463.1 & 70.9&PUCHEROS\\ 
6413.62 & -583.7 & 69.5&PUCHEROS\\ 
6421.61 & -611.8 & 63.3&PUCHEROS\\ 
6682.81 & -1879.2 & 73.0&PUCHEROS\\ 
6695.75 & -1550.4 & 63.8&PUCHEROS\\ 
6709.74 & -1658.5 & 67.8&PUCHEROS\\ 
6745.68 & -1808.4 & 61.8&PUCHEROS\\ 
6770.62 & -1745.6 & 64.5&PUCHEROS\\ 
6785.56 & -1894.7 & 61.4&PUCHEROS\\ 
6888.49 & 230.3 & 3.7&CHIRON\\ 
6904.47 & 178.8 & 3.8&CHIRON\\ 
6922.48 & 132.2 & 4.5&CHIRON\\ 
7022.82 & -107.6 & 4.5&CHIRON\\ 
7031.85 & -123.0 & 4.1&CHIRON\\ 
7040.83 & -143.3 & 4.0&CHIRON\\ 
7049.82 & -167.5 & 3.7  &CHIRON\\
\hline \vspace{-0.3cm} \\
\hline\hline
\end{tabular}
\end{table}

\begin{table}[]
\centering
  \caption{Radial velocity variations for HIP67890.}
  \label{tb1:instruments}
\begin{tabular}{rrrrr}
\hline\hline
JD      & RV         & err       & Instrument \\
-2450000& (ms$^{-1}$)  & (ms$^{-1}$) &      \\
\hline \vspace{-0.3cm} \\
3071.68 & 2347.3 & 40.0&FEROS\\
5317.64 & 369.7 & 5.1&FEROS \\
5379.65 & 284.3 & 4.4&FEROS \\
5428.53 & 254.4 & 5.1&FEROS \\
5729.64 & -39.0 & 11.1&FEROS \\
5729.64 & -43.1 & 10.2&FEROS \\
5744.61 & -53.9 & 5.3&FEROS \\
5744.61 & -57.8 & 5.0&FEROS \\
6047.63 & -210.4 & 5.7&FEROS \\
6047.63 & -216.4 & 6.4&FEROS \\
6056.61 & -235.1 & 4.9&FEROS \\
6066.63 & -239.6 & 5.1&FEROS \\
6099.61 & -227.2 & 4.4&FEROS \\
6110.59 & -247.0 & 4.3&FEROS \\
6140.59 & -238.6 & 4.0&FEROS \\
6140.60 & -241.7 & 3.8&FEROS \\
6321.80 & -223.1 & 6.8&FEROS \\
6321.80 & -222.5 & 6.9&FEROS \\
6342.79 & -202.6 & 7.6&FEROS \\
6342.79 & -197.0 & 7.0&FEROS \\
6412.71 & -178.3 & 3.2& FEROS\\
6431.65 & -182.5 & 5.4& FEROS\\
4868.25 & 874.1 & 1.3& UCLES\\
5227.21 & 439.0 & 1.4&UCLES\\
5380.95 & 241.6 & 1.6&UCLES\\
5580.26 & 15.9 & 1.4 &UCLES\\
5602.18 & 2.9 & 1.9&UCLES\\
6060.06 & -251.5 & 2.5&UCLES\\
6090.96 & -252.4 & 1.8&UCLES\\
6345.15 & -255.6 & 1.5&UCLES\\
6376.21 & -233.1 & 1.7&UCLES\\
6527.87 & -147.8 & 3.7&UCLES\\
\hline \vspace{-0.3cm} \\
\hline\hline
\end{tabular}
\end{table}

\begin{table}[]
\centering
  \caption{Radial velocity variations for HIP68099.}
  \label{tb1:instruments}
\begin{tabular}{rrrrr}
\hline\hline
JD      & RV         & err       & Instrument \\
-2450000& (ms$^{-1}$)  & (ms$^{-1}$) &      \\
\hline \vspace{-0.3cm} \\
5317.64 & -187.5 & 5.1&FEROS\\
5379.66 & -162.6 & 4.5&FEROS\\
5729.65 & -34.1 & 11.7&FEROS\\
5744.62 & -63.4 & 6.2&FEROS\\
6047.64 & -6.4 & 6.1&FEROS\\
6056.64 & -24.1 & 4.8&FEROS\\
6066.64 & -11.4 & 5.7&FEROS\\
6099.62 & 19.8 & 5.8&FEROS\\
6110.59 & 4.1 & 5.2&FEROS\\
6140.60 & 25.1 & 4.0&FEROS\\
6321.81 & 72.9 & 6.9&FEROS\\
6342.78 & 86.4 & 7.1&FEROS\\
6412.72 & 98.0 & 3.8&FEROS\\
6412.75 & 92.5 & 4.4&FEROS\\
6431.66 & 90.7 & 6.6&FEROS\\
6531.49 & -47.9 & 6.1&CHIRON\\
6654.84 & -58.0 & 4.6&CHIRON\\
6656.83 & -53.6 & 4.7&CHIRON\\
6673.87 & 1.7 & 7.0&CHIRON\\
6679.83 & 22.1 & 7.0&CHIRON\\
6695.88 & -13.9 & 5.3&CHIRON\\
6708.82 & -12.6 & 6.7&CHIRON\\
6722.70 & -9.7 & 5.5&CHIRON\\
6736.64 & -10.7 & 5.4&CHIRON\\
6752.69 & -29.0 & 5.1&CHIRON\\
6769.66 & -31.0 & 4.5&CHIRON\\
6790.54 & -28.0 & 5.4&CHIRON\\
6810.60 & 9.5 & 5.1&CHIRON\\
6833.58 & 28.8 & 4.9&CHIRON\\
6887.47 & -19.6 & 7.3&CHIRON\\
6908.47 & 2.5 & 4.5&CHIRON\\
7018.85 & 29.8 & 5.4&CHIRON\\
7025.85 & 25.4 & 4.4&CHIRON\\
7034.82 & 14.8 & 4.7&CHIRON\\
7050.79 & 48.0 & 4.7&CHIRON\\
7050.80 & 45.6 & 5.3&CHIRON\\
7061.78 & 46.6 & 5.0&CHIRON\\
7063.83 & 39.0 & 5.3 &CHIRON\\
\hline \vspace{-0.3cm} \\
\hline\hline
\end{tabular}
\end{table}

\begin{table}[]
\centering
  \caption{Radial velocity variations for HIP71778.}
  \label{tb1:instruments}
\begin{tabular}{rrrrr}
\hline\hline
JD      & RV         & err       & Instrument \\
-2450000& (ms$^{-1}$)  & (ms$^{-1}$) &      \\
\hline \vspace{-0.3cm} \\
5317.69 & -3148.7 & 7.9&FEROS\\
5379.67 & -2756.6 & 6.9&FEROS\\
5729.68 & -512.2 & 12.5&FEROS\\
5744.65 & -439.9 & 7.4&FEROS\\
6047.63 & 649.3 & 8.1&FEROS\\
6056.67 & 677.7 & 6.0&FEROS\\
6066.67 & 685.1 & 5.5&FEROS\\
6099.66 & 741.4 & 6.9&FEROS\\
6110.62 & 732.4 & 5.9&FEROS\\
6321.85 & 886.4 & 8.8&FEROS\\
6342.87 & 891.5 & 8.9&FEROS\\
6412.77 & 803.7 & 5.9&FEROS\\
6431.69 & 789.8 & 7.3&FEROS\\
\hline \vspace{-0.3cm} \\
\hline\hline
\end{tabular}
\end{table}

\begin{table}[]
\centering
  \caption{Radial velocity variations for HIP73758.}
  \label{tb1:instruments}
\begin{tabular}{rrrrr}
\hline\hline
JD      & RV         & err       & Instrument \\
-2450000& (ms$^{-1}$)  & (ms$^{-1}$) &      \\
\hline \vspace{-0.3cm} \\
4871.25 & 0.0 & 2.4&UCLES\\
5318.06 & -4836.2 & 1.2&UCLES\\
5971.20 & 14359.2 & 1.6&UCLES\\
6378.18 & 2138.4 & 2.0&UCLES\\
5317.71 & -6956.8 & 5.1&FEROS\\
5729.70 & -6061.2 & 10.3&FEROS\\
5744.67 & -2403.2 & 5.8&FEROS\\
5793.61 & 3287.7 & 5.1&FEROS\\
6047.69 & 1412.5 & 6.9&FEROS\\
6056.69 & 5293.0 & 5.0&FEROS\\
6066.69 & 11047.3 & 5.3&FEROS\\
6099.68 & -8026.5 & 5.2&FEROS\\
6110.66 & -7397.2 & 4.3&FEROS\\
6160.59 & 9045.7 & 3.2&FEROS\\
6321.86 & -3898.7 & 7.6&FEROS\\
6342.89 & 2985.1 & 7.2&FEROS\\
6412.63 & -5389.0 & 3.5&FEROS\\
6412.76 & -5368.4 & 3.9&FEROS\\
6431.72 & -119.0 & 6.1&FEROS\\
6565.49 & 12548.5 & 5.2&FEROS\\
\hline \vspace{-0.3cm} \\
\hline\hline
\end{tabular}
\end{table}

\begin{table}[]
\centering
  \caption{Radial velocity variations for HIP74188.}
  \label{tb1:instruments}
\begin{tabular}{rrrrr}
\hline\hline
JD      & RV         & err       & Instrument \\
-2450000& (ms$^{-1}$)  & (ms$^{-1}$) &      \\
\hline \vspace{-0.3cm} \\
5317.72 & 262.5 & 5.6&FEROS\\
5336.81 & 256.7 & 4.7&FEROS\\
5379.71 & 236.7 & 4.7&FEROS\\
5744.68 & 103.1 & 6.3&FEROS\\
5786.62 & 55.6 & 4.9&FEROS\\
5793.57 & 80.4 & 5.3&FEROS\\
6047.69 & -36.1 & 8.3&FEROS\\
6056.69 & -65.6 & 4.1&FEROS\\
6066.69 & -60.2 & 4.8&FEROS\\
6099.68 & -71.1 & 4.7&FEROS\\
6110.64 & -81.1 & 4.4&FEROS\\
6140.65 & -85.4 & 4.3&FEROS\\
6321.87 & -182.3 & 7.8 &FEROS\\
6342.89 & -193.4 & 7.6&FEROS\\
6412.79 & -219.9 & 4.5&FEROS\\
\hline \vspace{-0.3cm} \\
\hline\hline
\end{tabular}
\end{table}

\begin{table}[]
\centering
  \caption{Radial velocity variations for HIP75331.}
  \label{tb1:instruments}
\begin{tabular}{rrrrr}
\hline\hline
JD      & RV         & err       & Instrument \\
-2450000& (ms$^{-1}$)  & (ms$^{-1}$) &      \\
\hline \vspace{-0.3cm} \\
5317.73 & 555.3 & 6.0&FEROS\\
5379.72 & 523.6 & 4.3&FEROS\\
5729.71 & 290.5 & 10.5&FEROS\\
5744.69 & 277.0 & 6.2&FEROS\\
5793.59 & 230.4 & 5.4&FEROS\\
6047.71 & 1.9 & 7.2&FEROS\\
6056.72 & -33.7 & 5.0 &FEROS\\
6066.72 & -24.1 & 5.8 &FEROS\\
6099.71 & -51.3 & 4.4&FEROS\\
6110.67 & -68.8 & 4.7&FEROS\\
6160.58 & -116.7 & 4.0&FEROS\\
6321.89 & -305.5 & 6.8&FEROS\\
6342.92 & -308.9 & 7.2 &FEROS\\
6412.83 & -402.4 & 3.6  &FEROS\\
6565.48 & -567.3 & 6.4&FEROS\\
\hline \vspace{-0.3cm} \\
\hline\hline
\end{tabular}
\end{table}

\begin{table}[]
\centering
  \caption{Radial velocity variations for HIP76532.}
  \label{tb1:instruments}
\begin{tabular}{rrrrr}
\hline\hline
JD      & RV         & err       & Instrument \\
-2450000& (ms$^{-1}$)  & (ms$^{-1}$) &      \\
\hline \vspace{-0.3cm} \\
3070.79 & -4730.0 & 20.0&FEROS\\
5317.75 & 2842.4 & 5.1&FEROS\\
5336.85 & 2804.3 & 4.8&FEROS\\
5379.73 & 2647.9 & 6.5&FEROS\\
5428.60 & 2481.8 & 5.3&FEROS\\
5470.48 & 2287.6 & 12.3&FEROS\\
5729.76 & 980.0 & 10.5&FEROS\\
5729.76 & 979.4 & 11.3&FEROS\\
5744.70 & 888.1 & 6.1&FEROS\\
5744.71 & 894.7 & 5.8&FEROS\\
5793.61 & 599.6 & 4.3&FEROS\\
5793.62 & 605.3 & 4.0&FEROS\\
6047.72 & -888.1 & 6.4&FEROS\\
6047.72 & -887.5 & 6.8&FEROS\\
6099.71 & -1190.9 & 4.7&FEROS\\
6110.68 & -1252.2 & 3.1&FEROS\\
6412.85 & -2775.6 & 4.3&FEROS\\
6431.73 & -2869.8 & 5.5&FEROS\\
6565.49 & -3417.1 & 5.4&FEROS\\
5334.85 & 1932.0 & 15.7&FECH\\
5347.81 & 1873.8 & 12.5&FECH\\
5359.78 & 1837.2 & 10.9&FECH\\
5373.62 & 1808.1 & 11.2&FECH\\
5390.65 & 1778.8 & 14.6&FECH\\
5401.61 & 1705.8 & 17.3&FECH\\
5421.58 & 1627.0 & 13.1&FECH\\
5435.57 & 1579.3 & 15.1&FECH\\
5701.84 & 266.3 & 15.8&CHIRON\\
5704.84 & 260.7 & 16.2&CHIRON\\
5725.76 & 102.8 & 13.4&CHIRON\\
5742.69 & 0.3 & 13.2&CHIRON\\
5767.49 & -107.7 & 11.5&CHIRON\\
5782.62 & -196.6 & 7.2 &CHIRON\\
5804.56 & -326.8 & 7.0&CHIRON\\
6018.76 & -1607.6 & 7.6&CHIRON\\
6052.82 & -1797.6 & 5.8&CHIRON\\
6885.49 & -5341.8 & 5.6&CHIRON\\
6908.48 & -5394.0 & 5.7&CHIRON \\
\hline \vspace{-0.3cm} \\
\hline\hline
\end{tabular}
\end{table}

\begin{table}[]
\centering
  \caption{Radial velocity variations for HIP76569.}
  \label{tb1:instruments}
\begin{tabular}{rrrrr}
\hline\hline
JD      & RV         & err       & Instrument \\
-2450000& (ms$^{-1}$)  & (ms$^{-1}$) &      \\
\hline \vspace{-0.3cm} \\
3070.77 & -1487.9 & 20.0&FEROS\\
5470.48 & 3769.7 & 9.0&FEROS\\
5729.76 & 2758.5 & 10.1 &FEROS\\
5729.76 & 2748.3 & 11.9 &FEROS\\
5744.71 & 2700.9 & 7.0 &FEROS\\
5744.71 & 2695.1 & 6.6 &FEROS\\
6047.72 & 381.2 & 6.2 &FEROS\\
6047.72 & 381.5 & 6.9&FEROS\\
6066.72 & 254.4 & 6.3&FEROS\\
6066.72 & 257.1 & 6.5 &FEROS\\
6099.71 & 35.8 & 6.3 &FEROS\\
6140.66 & -323.6 & 4.5 &FEROS\\
6140.66 & -317.9 & 4.6 &FEROS\\
6160.60 & -434.4 & 4.9 &FEROS\\
6160.61 & -435.8 & 5.1   &FEROS\\
6412.80 & -1835.2 & 4.2  &FEROS\\
6431.73 & -1895.6 & 7.1 &FEROS\\
6565.49 & -2462.6 & 5.4 &FEROS\\
7072.89 & -3392.8 & 7.9 &FEROS\\
7072.89 & -3397.0 & 7.3&FEROS\\
5317.85 & 2058.7 & 13.9&FECH\\
5326.85 & 2097.8 & 17.7&FECH\\
5334.86 & 2058.5 & 17.1 &FECH\\
5359.78 & 2082.6 & 13.8 &FECH\\
5373.63 & 2111.7 & 14.2 &FECH\\
5390.66 & 2132.8 & 13.9 &FECH\\
5421.58 & 2063.5 & 21.1 &FECH\\
5435.58 & 2059.1 & 18.2 &FECH\\
5701.85 & 1187.7 & 15.8    &CHIRON\\
5704.85 & 1186.3 & 13.4 &CHIRON\\
5725.77 & 981.8 & 14.6 &CHIRON\\
5742.71 & 938.1 & 13.8 &CHIRON\\
5767.49 & 763.6 & 11.9 &CHIRON\\
5782.62 & 733.6 & 9.9 &CHIRON\\
5804.57 & 544.9 & 9.8 &CHIRON\\
6040.69 & -1294.1 & 7.1 &CHIRON\\
6050.69 & -1383.5 & 7.2   &CHIRON\\
6892.48 & -4973.9 & 7.8 &CHIRON\\
6918.48 & -5012.0 & 6.7 &CHIRON\\
7060.85 & -5155.3 & 6.8 &CHIRON\\
7091.78 & -5182.0 & 6.5   &CHIRON\\
\hline \vspace{-0.3cm} \\
\hline\hline
\end{tabular}
\end{table}

\clearpage
\begin{table}[]
\centering
  \caption{Radial velocity variations for HIP77888.}
  \label{tb1:instruments}
\begin{tabular}{rrrrr}
\hline\hline
JD      & RV           & err           & Instrument \\
-2450000& (ms$^{-1}$)  & (ms$^{-1}$) &      \\
\hline \vspace{-0.3cm} \\
5317.78 & -1181.8 & 6.7 &FEROS\\
5336.87 & -1130.7 & 5.7 &FEROS\\
5379.76 & -1036.5 & 5.9&FEROS\\
5428.61 & -950.4 & 5.7&FEROS\\
5470.49 & -831.8 & 10.9&FEROS\\
5729.77 & -216.3 & 11.7&FEROS\\
5744.72 & -165.6 & 5.7&FEROS\\
5793.63 & -59.2 & 4.2&FEROS\\
6047.73 & 532.2 & 7.2&FEROS\\
6099.73 & 654.9 & 6.0&FEROS\\
6110.69 & 695.9 & 5.7&FEROS\\
6140.67 & 721.1 & 4.3&FEROS\\
6412.66 & 1341.6 & 4.8&FEROS\\
6565.51 & 1626.8 & 6.8&FEROS\\
5319.11 & -1828.3 & 1.4&UCLES\\
5707.09 & -919.3 & 2.3&UCLES\\
6090.13 & 0.0 & 2.9&UCLES\\
6377.16 & 614.7 & 2.7&UCLES\\
6526.95 & 938.3 & 4.3&UCLES\\
6746.27 & 1270.9 & 1.8&UCLES\\
\hline \vspace{-0.3cm} \\
\hline\hline
\end{tabular}
\end{table}

\begin{table}[]
\centering
  \caption{Radial velocity variations for HIP83224.}
  \label{tb1:instruments}
\begin{tabular}{rrrrr}
\hline\hline
JD      & RV         & err       & Instrument \\
-2450000& (ms$^{-1}$)  & (ms$^{-1}$) &      \\
\hline \vspace{-0.3cm} \\
5319.17 & -0.2 & 0.6&UCLES\\
5382.06 & 10851.2 & 0.9&UCLES\\
5379.77 & 7098.4 & 5.0&FEROS\\ 
5428.63 & -1423.7 & 4.5&FEROS\\
5457.56 & -8124.6 & 4.6&FEROS\\
5470.54 & -6797.2 & 9.5 &FEROS\\
5744.75 & 9187.8 & 4.8   &FEROS\\
5786.64 & -6707.6 & 4.9 &FEROS\\
5793.65 & -8008.5 & 5.8 &FEROS\\
6047.76 & 2769.6 & 7.7 &FEROS\\
6066.75 & 6081.5 & 6.4&FEROS\\
6160.63 & -7192.9 & 4.1  &FEROS\\
6412.84 & 5981.0 & 5.4 &FEROS\\
6431.75 & 8740.5 & 7.3 &FEROS\\
6472.76 & -3940.8 & 5.5 &FEROS\\
6565.52 & 2336.6 & 4.8 &FEROS\\
6561.49 & -17661.9 & 90.8&PUCHEROS\\
6582.49 & -14280.6 & 78.3&PUCHEROS\\
6709.84 & -22002.3 & 191.8&PUCHEROS\\
6745.79 & -15527.5 & 80.5&PUCHEROS\\
6770.81 & -11310.5 & 71.6&PUCHEROS\\
6785.77 & -9943.8 & 72.0 &PUCHEROS\\
\hline \vspace{-0.3cm} \\
\hline\hline
\end{tabular}
\end{table}

\begin{table}[]
\centering
  \caption{Radial velocity variations for HIP101911.}
  \label{tb1:instruments}
\begin{tabular}{rrrrr}
\hline\hline
JD      & RV         & err       & Instrument \\
-2450000& (ms$^{-1}$)  & (ms$^{-1}$)  & -     \\
\hline \vspace{-0.3cm} \\
5137.94 & -228.8 & 1.9&UCLES\\
5381.22 & -175.1 & 1.4&UCLES\\
6052.22 & -5.8 & 1.7&UCLES\\
6090.17 & 0.0 & 1.6&UCLES\\
6470.24 & 153.4 & 2.2&UCLES\\
6527.11 & 173.9 & 4.7&UCLES\\
5317.83 & -125.7 & 3.6&FEROS\\
5336.92 & -103.9 & 4.4&FEROS\\
5366.89 & -102.2 & 5.2&FEROS\\
5379.85 & -112.2 & 3.6&FEROS\\
5428.72 & -115.0 & 3.8&FEROS\\
5457.64 & -92.8 & 3.4&FEROS\\
5470.63 & -80.0 & 8.2&FEROS\\
5786.81 & -19.4 & 4.6&FEROS\\
5793.77 & -14.7 & 6.0&FEROS\\
5793.77 & -14.1 & 6.5&FEROS\\
6056.79 & 68.8 & 4.3&FEROS\\
6066.79 & 65.9 & 5.9&FEROS\\
6110.79 & 88.1 & 4.8&FEROS\\
6160.72 & 90.2 & 3.2&FEROS\\
6241.52 & 110.6 & 6.5&FEROS\\
6251.53 & 111.2 & 4.0&FEROS\\
6565.59 & 245.2 & 5.1&FEROS\\
6515.76 & -118.9 & 7.1&CHIRON\\
6524.64 & -113.8 & 7.0&CHIRON\\
6530.64 & -99.4 & 7.2&CHIRON\\
6539.70 & -90.7 & 31.7&CHIRON\\
6895.59 & 57.0 & 6.5&CHIRON\\
6910.54 & 61.3 & 6.5&CHIRON\\
6918.57 & 60.0 & 6.5&CHIRON\\
6924.57 & 55.5 & 5.9&CHIRON\\
6926.59 & 49.1 & 6.2&CHIRON\\
6940.57 & 60.6 & 7.0&CHIRON\\
6950.54 & 79.3 & 6.5&CHIRON\\
\hline \vspace{-0.3cm} \\
\hline\hline
\end{tabular}
\end{table}

\begin{table}[]
\centering
  \caption{Radial velocity variations for HIP103836.}
  \label{tb1:instruments}
\begin{tabular}{rrrrr}
\hline\hline
JD      & RV         & err       & Instrument \\
-2450000& (ms$^{-1}$)  & (ms$^{-1}$) &      \\
\hline \vspace{-0.3cm} \\
5074.03 & -177.1 & 1.9&UCLES\\ 
5381.17 & -121.2 & 1.4&UCLES\\ 
5456.03 & -104.5 & 1.9&UCLES\\ 
5707.29 & -54.4 & 1.2&UCLES\\ 
5760.08 & -28.9 & 1.6&UCLES\\ 
5783.19 & -57.5 & 2.6&UCLES\\ 
5841.97 & -41.7 & 1.5&UCLES\\ 
6052.26 & 20.0 & 1.5&UCLES\\ 
6060.25 & 17.8 & 2.3&UCLES\\ 
6089.26 & 0.0 & 1.6&UCLES\\ 
6469.16 & 82.0 & 1.8&UCLES\\ 
6494.08 & 84.8 & 1.7&UCLES\\ 
6529.05 & 107.2 & 1.6&UCLES\\ 
6747.29 & 182.1 & 1.4 &UCLES\\
5326.89 & -156.6 & 10.0 &FECH\\
5338.85 & -173.5 & 19.2&FECH\\ 
5347.86 & -131.8 & 10.9&FECH\\ 
5373.72 & -179.2 & 10.7&FECH\\ 
5390.74 & -127.3 & 11.2&FECH\\ 
5401.74 & -139.6 & 17.7&FECH\\ 
5517.54 & -123.8 & 13.2&FECH\\ 
5531.54 & -109.7 & 12.0&FECH\\ 
5705.83 & -103.1 & 10.6&CHIRON\\ 
5709.82 & -62.9 & 24.2&CHIRON\\ 
5725.86 & -105.3 & 9.4&CHIRON\\ 
5737.89 & -113.3 & 12.2&CHIRON\\ 
5756.79 & -90.5 & 10.8&CHIRON\\ 
5767.78 & -107.3 & 10.3&CHIRON\\ 
5786.77 & -72.5 & 6.6&CHIRON\\ 
5798.73 & -80.3 & 6.4&CHIRON\\ 
6223.51 & 7.7 & 7.7&CHIRON\\ 
6230.53 & -1.7 & 7.2&CHIRON\\ 
6241.56 & 22.8 & 7.0&CHIRON\\ 
6250.56 & 38.1 & 6.6&CHIRON\\ 
6464.98 & 31.3 & 6.3&CHIRON\\ 
6480.72 & 38.9 & 5.8&CHIRON\\ 
6508.83 & 66.1 & 8.5&CHIRON\\ 
6515.77 & 68.8 & 6.1&CHIRON\\ 
6531.76 & 62.0 & 5.5&CHIRON\\ 
6560.69 & 79.8 & 6.7&CHIRON\\ 
6571.52 & 82.0 & 6.7&CHIRON\\ 
6896.64 & 152.8 & 5.1&CHIRON\\ 
6910.54 & 163.9 & 4.9&CHIRON\\ 
6911.57 & 143.6 & 5.2&CHIRON\\ 
6921.69 & 139.9 & 4.9&CHIRON\\ 
6924.59 & 167.2 & 5.0&CHIRON\\ 
6931.55 & 151.3 & 5.4&CHIRON\\ 
6938.58 & 151.6 & 5.4&CHIRON\\ 
6976.52 & 156.9 & 4.9&CHIRON\\ 
6976.52 & 153.8 & 4.9  &CHIRON\\ 
\hline \vspace{-0.3cm} \\
\hline\hline
\end{tabular}
\end{table}

\begin{table}[]
\centering
  \caption{Radial velocity variations for HIP104148.}
  \label{tb1:instruments}
\begin{tabular}{rrrrr}
\hline\hline
JD      & RV         & err       & Instrument \\
-2450000& (ms$^{-1}$)  & (ms$^{-1}$) &      \\
\hline \vspace{-0.3cm} \\
5317.83 & -4415.9 & 3.3&FEROS\\  
5336.92 & -4696.3 & 4.2&FEROS\\ 
5366.92 & -5069.6 & 5.4&FEROS\\
5379.88 & -5178.5 & 3.4&FEROS\\
5428.77 & -5411.3 & 4.6&FEROS\\
5457.647 & -5377.7 & 3.5&FEROS\\
5470.64 & -5280.0 & 8.1&FEROS\\
5729.85 & 65.4 & 17.7&FEROS\\
5786.79 & 1401.2 & 4.1&FEROS\\
5786.79 & 1398.4 & 3.8&FEROS\\
5793.81 & 1581.0 & 4.5&FEROS\\
5793.81 & 1583.3 & 3.7&FEROS\\
6056.80 & 5333.7 & 3.9&FEROS\\
6110.79 & 5579.6 & 3.4&FEROS\\
6160.72 & 5722.2 & 2.7&FEROS\\
6251.55 & 5661.8 & 4.1&FEROS\\
6431.93 & 4461.7 & 5.8&FEROS\\
6565.59 & 2641.0 & 5.2&FEROS\\
\hline \vspace{-0.3cm} \\
\hline\hline
\end{tabular}
\end{table}

\begin{table}[]
\centering
  \caption{Radial velocity variations for HIP106055.}
  \label{tb1:instruments}
\begin{tabular}{rrrrr}
\hline\hline
JD      & RV         & err       & Instrument \\
-2450000& (ms$^{-1}$)  & (ms$^{-1}$) &      \\
\hline \vspace{-0.3cm} \\
5317.87 & -1528.8 & 4.2&FEROS\\ 
5366.90 & -1272.0 & 5.3&FEROS\\
5379.86 & -1209.3 & 4.8&FEROS\\
5428.74 & -977.6 & 4.1&FEROS\\
5457.67 & -869.2 & 2.6&FEROS\\
5470.65 & -801.9 & 7.8&FEROS\\
5744.84 & 277.5 & 5.3&FEROS\\
5793.80 & 384.6 & 4.6&FEROS\\
6047.84 & 908.1 & 6.2&FEROS\\
6110.81 & 989.8 & 4.8&FEROS\\
6160.74 & 1040.5 & 5.4&FEROS\\
6241.54 & 1119.4 & 7.2&FEROS\\
6251.56 & 1115.6 & 6.3&FEROS\\
6565.60 & 823.2 & 5.4&FEROS\\
\hline \vspace{-0.3cm} \\
\hline\hline
\end{tabular}
\end{table}

\begin{table}[]
\centering
  \caption{Radial velocity variations for HIP107122.}
  \label{tb1:instruments}
\begin{tabular}{rrrrr}
\hline\hline
JD      & RV         & err       & Instrument \\
-2450000& (ms$^{-1})$  &(ms$^{-1})$ &      \\
\hline \vspace{-0.3cm} \\
5317.85 & -264.3 & 5.0&FEROS\\ 
5366.935 & -217.5 & 5.2&FEROS\\ 
5379.89 & -200.7 & 4.8&FEROS\\ 
5428.78 & -141.9 & 5.4&FEROS\\ 
5457.68 & -119.0 & 3.5&FEROS\\ 
5470.67 & -108.4 & 7.4&FEROS\\ 
5744.85 & -46.4 & 5.2&FEROS\\ 
5786.83 & -17.2 & 4.2&FEROS\\ 
5793.83 & -10.2 & 4.0&FEROS\\ 
6056.81 & 53.7 & 6.9&FEROS\\ 
6110.85 & 63.4 & 4.0&FEROS\\ 
6160.75 & 110.5 & 3.4&FEROS\\ 
6241.55 & 139.8 & 7.6&FEROS\\ 
6251.58 & 137.4 & 4.6&FEROS\\ 
6431.94 & 146.8 & 6.5&FEROS\\ 
6565.57 & 236.3 & 5.8&FEROS\\ 
6565.68 & 237.6 & 5.6&FEROS\\ 
\hline \vspace{-0.3cm} \\
\hline\hline
\end{tabular}
\end{table}

\section{Atmospheric Parameters}
\begin{table}[]
\centering
  \caption{Atmospheric parameters for Setiawan data}
  \label{tb:Atmospheric}
\begin{tabular}{lrccc}
\hline\hline
Object & [Fe/H] & T   & $\log g$     & V$_t$ \\
       & (dex)  & (K) &(cm/s$^2$)  & (kms$^{-1}$)   \\
\hline \vspace{-0.3cm} \\
HD101321 & -0.11  & 4871.0 & 3.05 & 1.08 \\
HD10700  & -0.49  & 5454.0 & 4.69 & 0.55 \\
HD107446 & -0.12  & 4337.0 & 1.84 & 1.57 \\
HD10761  &  0.08  & 5116.0 & 2.83 & 1.54 \\
HD108570 & -0.02  & 5057.0 & 3.54 & 0.96 \\
HD110014 &  0.30  & 4805.0 & 2.98 & 1.90 \\
HD111884 &  0.01  & 4442.0 & 2.31 & 1.39 \\
HD113226 &  0.25  & 5236.0 & 3.07 & 1.47 \\
HD115439 & -0.15  & 4358.0 & 1.91 & 1.37 \\
HD115478 &  0.12  & 4463.0 & 2.33 & 1.33 \\
HD11977  & -0.14  & 5032.0 & 2.72 & 1.33 \\
HD121416 &  0.12  & 4692.0 & 2.52 & 1.41 \\
HD122430 &  0.03  & 4485.0 & 2.12 & 1.44 \\
HD12438  & -0.55  & 5046.0 & 2.56 & 1.46 \\
HD124882 & -0.23  & 4388.0 & 1.81 & 1.46 \\
HD125560 &  0.28  & 4597.0 & 2.47 & 1.33 \\
HD131109 &  0.06  & 4314.0 & 2.01 & 1.39 \\
HD131977 &  0.01  & 4842.0 & 4.53 & 0.77 \\
HD136014 & -0.39  & 4983.0 & 2.57 & 1.51 \\
HD148760 &  0.20  & 4782.0 & 2.88 & 1.11 \\
HD151249 & -0.19  & 4289.0 & 1.79 & 1.54 \\
HD152334 &  0.14  & 4392.0 & 2.22 & 1.40 \\
HD152980 &  0.06  & 4367.0 & 1.99 & 1.66 \\
HD156111 & -0.29  & 5235.0 & 3.96 & 0.60 \\
HD159194 &  0.29  & 4639.0 & 2.59 & 1.31 \\
HD16417  &  0.16  & 5861.0 & 4.14 & 1.03 \\
HD165760 &  0.10  & 5066.0 & 2.81 & 1.40 \\
HD169370 & -0.12  & 4659.0 & 2.62 & 1.21 \\
HD174295 & -0.14  & 5000.0 & 2.84& 1.35 \\
HD175751 &  0.02  & 4750.0 & 2.66 & 1.42 \\
HD176578 &  0.02  & 4977.0 & 3.45 & 1.04 \\
HD177389 & -0.0   & 5139.0 & 3.42 & 1.11 \\
HD179799 & -0.0   & 4928.0 & 3.29 & 1.09 \\
HD18322  &  0.02  & 4727.0 & 2.70 & 1.24 \\
HD187195 &  0.12  & 4577.0 & 2.70 & 1.37 \\
HD18885  &  0.20  & 4823.0 & 2.75 & 1.39 \\
HD18907  & -0.5   & 5181.0 & 3.72 & 0.89 \\
HD189319 & -0.2   & 4195.0 & 1.71 & 1.66 \\
HD190608 &  0.08  & 4803.0 & 3.08 & 1.17 \\
HD197635 & -0.0   & 4677.0 & 2.50 & 1.49 \\
HD198232 &  0.15  & 5063.0 & 2.92 & 1.41 \\
HD199665 &  0.09  & 5138.0 & 3.17 & 1.22 \\
HD21120  & -0.0   & 5230.0 & 2.50 & 1.63 \\
HD2114   &  0.05  & 5324.0 & 2.66 & 1.64 \\
HD2151   & -0.0   & 5850.0 & 3.98 & 1.13 \\
HD218527 & -0.1   & 5094.0 & 2.97 & 1.44 \\
HD219615 & -0.4   & 5000.0 & 2.64 & 1.46 \\
HD224533 &  0.04  & 5119.0 & 2.95 & 1.39 \\
HD22663  &  0.05  & 4594.0 & 2.42 & 1.08 \\
HD23319  &  0.39  & 4735.0 & 2.85 & 1.34 \\
HD23940  & -0.28  & 4914.0 & 2.68 & 1.42 \\
HD26923  &  0.05  & 6090.0 & 4.49 & 1.06 \\
HD27256  &  0.18  & 5266.0 & 2.89 & 1.49 \\
HD27371  &  0.23  & 5094.0 & 3.07 & 1.47 \\
HD27697  &  0.25  & 5140.0 & 3.01 & 1.44 \\
HD32887  &  0.01  & 4409.0 & 2.06 & 1.64 \\
HD34642  &  0.00  & 4908.0 & 3.19 & 0.99 \\
\hline \vspace{-0.3cm} \\
\hline\hline
\end{tabular}
\end{table}

\begin{table}[]
\centering
  \caption{Continued Table\,\ref{tb:Atmospheric}}
  \label{tbB:AtmosfericB}
\begin{tabular}{lrccc}
\hline\hline
Object & [Fe/H] & T   & $\log g$     & V$_t$ \\
       & (dex)  & (K) &(cm/s$^2$)  & (kms$^{-1}$)   \\
\hline \vspace{-0.3cm} \\
HD36189  &  0.03  & 5072.0 & 2.34 & 1.67 \\
HD36848  &  0.45  & 4723.0 & 2.93 & 1.12 \\
HD40176  &  0.39  & 4899.0 & 2.75 & 1.45 \\
HD47205  &  0.23  & 4821.0 & 3.09 & 1.09 \\
HD47536  & -0.70  & 4414.0 & 1.93 & 1.44 \\
HD50778  & -0.29  & 4250.0 & 1.67 & 1.49 \\
HD61935  &  0.06  & 4924.0 & 2.77 & 1.38 \\
HD62644  &  0.08  & 5546.0 & 3.98 & 0.93 \\
HD62902  &  0.37  & 4516.0 & 2.73 & 1.39 \\
HD62902  &  0.37  & 4516.0 & 2.73 & 1.39 \\
HD63697  &  0.26  & 4576.0 & 2.59 & 1.35 \\
HD65695  & -0.11  & 4560.0 & 2.27 & 1.35 \\
HD65735  & -0.04  & 4693.0 & 2.51 & 1.47 \\
HD70982  &  0.07  & 5163.0 & 3.01 & 1.35 \\
HD72650  &  0.12  & 4440.0 & 2.28 & 1.36 \\
HD7672   & -0.33  & 5107.0 & 2.92 & 1.39 \\
HD78647  & -0.03  & 4403.0 & 1.89 & 2.61 \\
HD81361  &  0.25  & 5032.0 & 3.32 & 1.24 \\
HD81797  &  0.08  & 4307.0 & 1.88 & 1.75 \\
HD83441  &  0.14  & 4766.0 & 2.74 & 1.27 \\
HD85035  &  0.11  & 4808.0 & 3.24 & 1.09 \\
HD90957  &  0.24  & 4460.0 & 2.45 & 1.44 \\
HD92588  &  0.08  & 5235.0 & 3.89 & 0.91 \\
HD93257  &  0.27  & 4719.0 & 3.02 & 1.14 \\
HD93773  &  0.00  & 5124.0 & 3.07 & 1.43 \\
HD99167  &  0.05  & 4624.0 & 2.75 & 1.62 \\
\hline \vspace{-0.3cm} \\
\hline\hline
\end{tabular}
\end{table}

\end{appendix}
\end{document}